\documentclass[fleqn,usenatbib]{mnras}
\makeatletter
\def\endfigure{\end@float}
\def\endfigure{\end@float}
\def\endtable{\end@float}
\def\endtable{\end@float}
\makeatother

\usepackage{mathptmx}

\usepackage[T1]{fontenc}

\usepackage[british]{babel}             
   
\usepackage{graphicx}	
\usepackage{amsmath}	
\usepackage{amssymb}	

\usepackage[bottom]{footmisc}
\usepackage{footnote}
\makesavenoteenv{tabular}

\usepackage{tabularx}

\usepackage{hyperref}	
\hypersetup{colorlinks=true,linkcolor=blue,citecolor=blue,filecolor=blue,urlcolor=blue}

\usepackage[caption=false]{subfig}
\usepackage{booktabs}
\usepackage{pbox}
\usepackage{multirow}
\usepackage{multicol}

\newcommand{\tool}{\texttt{CAESAR}}
\newcommand{\toolinfo}{\emph{Compact And Extended Source Automated Recognition}}
\newcommand{\RMS}{\textrm{rms}}

\title[Extended source detection in the SCORPIO survey]{Automated detection of extended sources in radio maps: progress from the SCORPIO survey}

\author[S. Riggi et al.]{
S. Riggi$^{1}$\thanks{E-mail: sriggi@oact.inaf.it (SR)},
A. Ingallinera$^{1}$,
P. Leto$^{1}$,
F. Cavallaro$^{1,2,3}$,
F. Bufano$^{1}$,
F. Schillir\`{o}$^{1}$,
\newauthor
C. Trigilio$^{1}$,
G. Umana$^{1}$,
C. S. Buemi$^{1}$
and R. P. Norris$^{3,4}$
\\
$^{1}$INAF- Osservatorio Astrofisico di Catania, Via S. Sofia 78, I-95123, Catania, Italy\\
$^{2}$Universit\`{a} di Catania, Dipartimento di Fisica e Astronomia, Via Santa Sofia, 64, I-95123 Catania, Italy\\
$^{3}$CSIRO Astronomy and Space Science, PO Box 76, Epping, NSW 1710, Australia\\
$^{4}$Western Sidney University, Locked Bag 1797, Penrith South, NSW 1797, Australia
}

\date{Accepted 2016 April 22. Received 2016 April 22; in original form 2016 January 13}

\pubyear{2016}

\begin{document}
\label{firstpage}
\pagerange{\pageref{firstpage}--\pageref{lastpage}}
\maketitle

\begin{abstract}
Automated source extraction and parameterization represents a crucial challenge for the next-generation radio interferometer surveys, such as those performed with 
the Square Kilometre Array (SKA) and its precursors. In this paper we present a new algorithm, dubbed \tool{} (\toolinfo{}), to detect and parametrize extended sources in radio interferometric maps. 
It is based on a pre-filtering stage, allowing image denoising, compact source suppression and enhancement of diffuse emission, followed by an adaptive superpixel 
clustering stage for final source segmentation. A parameterization stage provides source flux information and a wide range of morphology estimators for post-processing 
analysis. We developed \tool{} in a modular software library, including also different methods for 
local background estimation and image filtering, along with alternative algorithms for both compact and diffuse source extraction.
The method was applied to real radio continuum data collected at the Australian Telescope Compact Array (ATCA) within the SCORPIO project, a pathfinder 
of the ASKAP-EMU survey. The source reconstruction capabilities were studied over different test fields in the presence of compact sources, imaging artefacts and diffuse 
emission from the Galactic plane and compared with existing algorithms. When compared to a human-driven analysis, the designed algorithm was found capable of detecting known target sources 
and regions of diffuse emission, outperforming alternative approaches over the considered fields.
\end{abstract}




\section{Introduction}\label{IntroductionSection}
A new era in radio astronomy is approaching with the upcoming continuum surveys \citep{Norris2013} planned at the SKA precursors telescopes, such as the 
\textit{Westerbork Observations of the Deep APERTIF Northern-Sky} (WODAN) \citep{Rottgering2010} 
at the Westerbork Synthesis Radio Telescope (WSRT), the \textit{Evolutionary Map of the Universe} (EMU) survey \citep{Norris2011} at the ASKAP array and 
the \textit{MeerKAT International GigaHertz Tiered Extragalactic Exploration} (MIGHTEE) survey \citep{Van der Heyden2010} at the Meerkat observatory. A considerable improvement 
is expected in sensitivity, resolution and instantaneous field of view compared to previous surveys. For instance, WODAN and EMU will jointly provide full sky 
coverage at 1.3 GHz with an unprecedented sensitivity down to 10-15 $\mu$Jy/beam and resolution around 10-15 arcsec. Phased Array Feed (PAF) technology will allow 
instantaneous field of view of 8 and 30 deg$^{2}$ for WODAN-APERTIF and ASKAP respectively and a corresponding increase in survey speed of a factor $\sim$20 with respect 
to VLA. MIGHTEE will allow even better sensitivities (0.1-1 $\mu$Jy/beam rms) although with a reduced field of view (1 deg$^{2}$). A dramatic gain in sensitivity (a factor 100) and 
field of view will be achieved with the future operations of the SKA.

New challenges are expected to be brought by these significant advances. One is related to the data product throughput (e.g. spectral-imaging data cubes) expected to be generated 
by the SKA precursor telescopes, ranging from tens of gigabytes to several petabytes\footnote{ASKAP is expected to generate several petabytes per year of HI cube.}, and by the future SKA observatory, of the order of hundreds of terabytes per data 
cube in SKA1 and one order of magnitude higher in SKA Phase II \citep{Kitaeff2015}. For instance, up to 3 exabytes of fully processed data are expected in one 
year of full SKA1 operation \citep{Alexander2009}. Such amount of data cannot be processed nor stored and visualized on local computing resources, at least using conventional 
data formats so far used in astronomy. 

Furthermore, with the increase in sensitivity and surveyed sky area, a population of millions of sources will be potentially detectable making human-driven source extraction 
unfeasible. For example, the EMU survey is expected to generate a catalogue of $\sim$70 millions of sources detected at the 5$\sigma$ level of 50 $\mu$Jy/beam \citep{Norris2011}.

For these reasons considerable efforts are currently focused on the development of algorithms to process imaging data and extract sources in a fast and mostly automated 
way and, at the same time, on the search for new data standards and image compression formats (e.g. see \citealt{Kitaeff2014}).

While extensive studies have been performed on compact source search with several algorithms developed \citep{Hancock2012,Whiting2009,Whiting2012,Hopkins2002,Bertin1996,Hales2012,
Peracaula2015,Hopkins2015}, particularly in 
the context of the ASKAP telescope, detection of extended sources in a completely unsupervised way (e.g. without requiring any a priori information or source 
templates) is still a partially explored field, at least for the radio domain. This motivates investing resources on exploring completely new methods or re-adapting known 
algorithms to the radio imaging case.

Different approaches have been recently proposed in such direction. Some of them make use of conventional thresholding methods in the image wavelet or curvelet domain 
(e.g. see \citealt{Peracaula2011}), others employ compressive sampling techniques (e.g. \citealt{Dabbech2015}). Other studies employ the Circle Hough transform to detect circular-like objects, such as supernova remnants or bent-tail
radio galaxies \citep{Hollitt2012}. In \citet{Norris2011} several methods from the Computer Vision domain have been reviewed. Waterfalling segmentation, circular or elliptical 
Hough transform and region growing were indicated as the most suited to the problem of extended source search.

In the context of the SCORPIO project \citep{Umana2015} (hereafter denoted as "Paper I", see Section~\ref{SCORPIOProjectSection}), a pathfinder of the ASKAP EMU survey, and in view of 
the next-generation SKA surveys, we started to develop algorithms for automated source detection and classification. The designed method exploits some of the techniques and algorithms already 
in use in other source finders, aiming to combine their best features, but also introduces new features, particularly on the background estimation, detection of 
extended sources and source parameterization. We will therefore focus on these novel aspects throughout the paper. A description of the method, based on a superpixel segmentation 
and hierarchical merging, is presented in Section~\ref{MethodSection}. The algorithm has been tested on SCORPIO real radio data observed at the ATCA array down to a 
sensitivity of 30 $\mu$Jy/beam. Typical results achieved on sample field scenarios are presented and discussed in Section~\ref{ResultSection}, along with tests performed 
on the same fields observed at different wavelengths.

\begin{figure*}
\centering%
\includegraphics[scale=0.55]{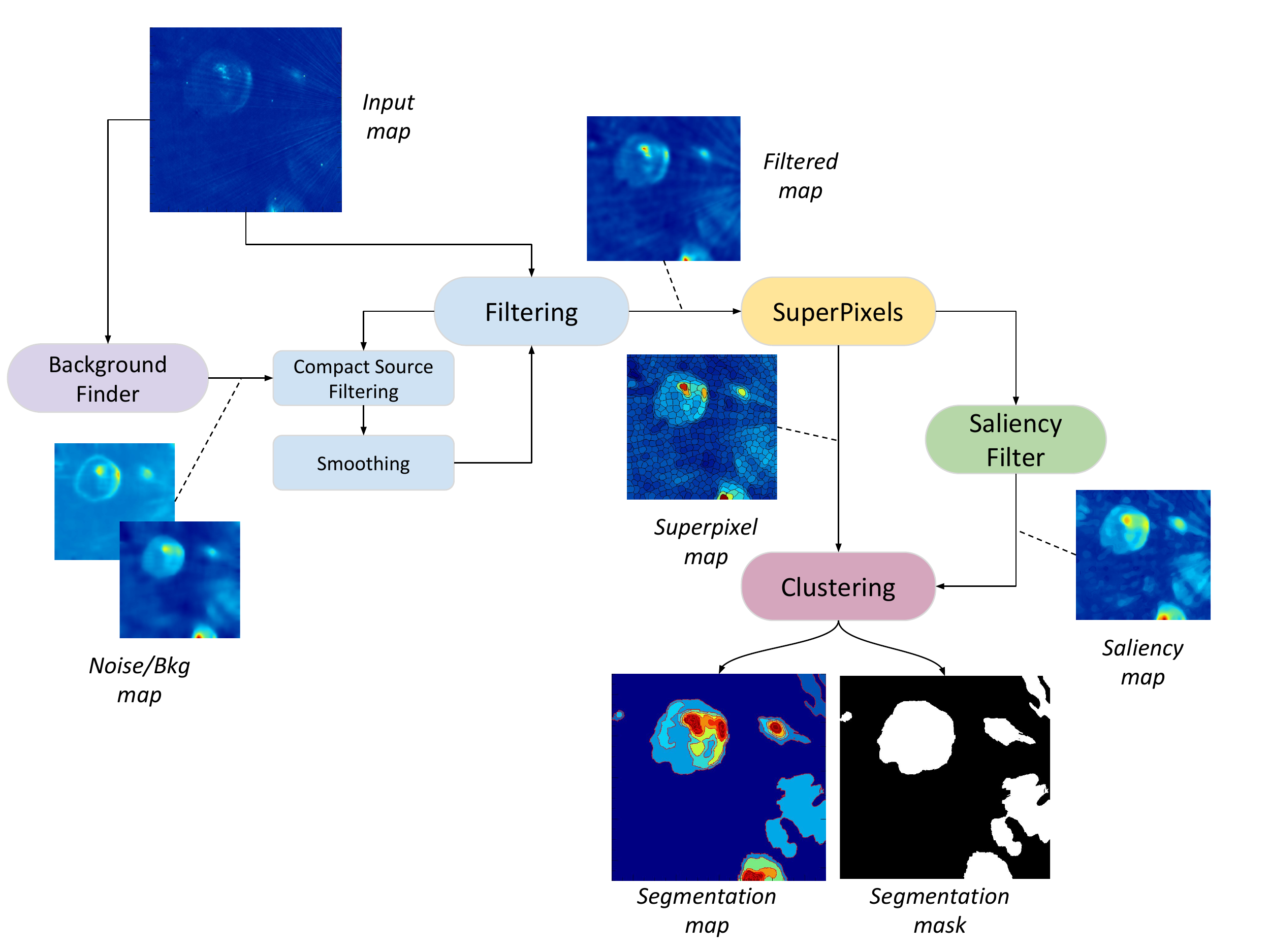}
\caption{Schematic pipeline of the designed source finder algorithm.}
\label{SourceFinderPipelineFig}
\end{figure*}

\section{The SCORPIO project}\label{SCORPIOProjectSection}
The SCORPIO project is a blind deep radio survey of a $2\times2$ deg$^2$ sky patch toward the Galactic plane, using the ATCA array in several 
configurations. The survey has been conducted at 2.1 GHz between 2011 and 2015 and achieved an average resolution around 10 arcsec. Further observations are already scheduled in 2016.
The major scientific goals of the SCORPIO project is to search for different populations of Galactic radio point sources and the study of circumstellar envelopes 
(related to young or evolved massive stars, planetary nebulae and supernova remnants) which is extremely important for understanding the Galaxy evolution (e.g. ISM
chemical enrichment, star formation triggering, etc.). Besides these scientific outputs, SCORPIO will be used
as a test-bed for the EMU survey, guiding its design strategy for the Galactic plane sections. In particular, this includes exploring suitable strategies for effectively imaging and 
extracting sources embedded in the diffuse emission expected at low Galactic latitudes and investigating to what extent they can be employed in the EMU survey.

The SCORPIO observations have produced a radio mosaic map of 133 single pointings with an \RMS{} down to 30 $\mu$Jy/beam. A pixel size of 1.5" is chosen for the final map. 
This sensitivity and a good $uv$-plane coverage have allowed the discovery of about 1000 new faint radio point sources and to satisfactorily map tens of extended sources. 
Preliminary results on a smaller pilot region of the SCORPIO field have already been published in Paper I, while the complete data reduction and analysis is 
still in progress.

\section{A segmentation method for extended source detection}\label{MethodSection}
Detection of extended sources represents a hard task for source finder algorithms. The main difficulties are due to the intrinsic emission pattern, which is usually fainter compared 
to compact sources (e.g. below the conventional 5$\sigma$ significance level) and spread over disjointed areas (e.g. unlike the adjacency assumption taken in compact source finders). 
In addition, object borders are usually soft thus the standard edge detector algorithms are not fully sensitive to them. 
Spatial filters are therefore often employed to enhance the emission at some given scale.

Another issue is related to the estimation of reliable significance levels for detection. In fact, the widely used method for local noise and background estimation is typically 
biased around extended source regions, namely higher significance levels are artificially imposed for detection with respect to other image regions, free of diffuse emission. Under these 
conditions the source is likely to be undetected particularly if it has a large extension.

Ideally, the source extraction task should provide a two-level hierarchical information: a segmentation of the input map into background and foreground regions associated with 
a source object, and, for each of the them, a collection of nested regions representing source features (e.g. clumps, shells, blobs) also at 
different scales. 

To this goal, we designed a multi-stage method based on image superpixel generation and hierarchical clustering. 
A schematic pipeline of the algorithm stages is shown in Fig.~\ref{SourceFinderPipelineFig} and summarized below:
\begin{enumerate}
 \item \emph{Filtering}: To enhance extended structures, bright compact sources need to be filtered out from the map and a residual image generated and used as 
 input for the following stages. Compact source extraction, discussed with more details in Section~\ref{CompactSourceFindingSect}, requires the computation of the background 
 and noise maps to threshold the image at a suitable significance level.
 
 Furthermore, a smoothing stage is introduced on the residual image to suppress texture-like features due to imaging artefacts around the brightest sources and to source residuals left
 after the previous dilation stage. An edge-preserving guided filter \citep{He2013} was found to provide optimal performances among the tested filters.
 \item \emph{Extended source extraction}: The smoothed residual image is used as input for the segmentation algorithm described in Section \ref{SegmentationAlgorithmSection}. 
 It consists of three main stages: firstly, an over-segmentation of the image into a collection of superpixels or regions is generated and a set of appearance parameters (both intensity- and 
 spatial-based) computed for each region; then, a saliency map is computed in the second stage from region dissimilarities and used to drive region merging at the third stage,  
 which is a sequence of clustering steps producing a collection of segmented regions or a binary mask as the final output. 
 \item \emph{Source parametrization}: A set of morphological parameters is calculated over the segmented regions and delivered to the user.
\end{enumerate}
Additional details concerning each algorithm step are given in the following sections.
 
\begin{figure*}
\subfloat[Field A\label{ScorpioFieldFig1}]{\includegraphics[scale=0.28]{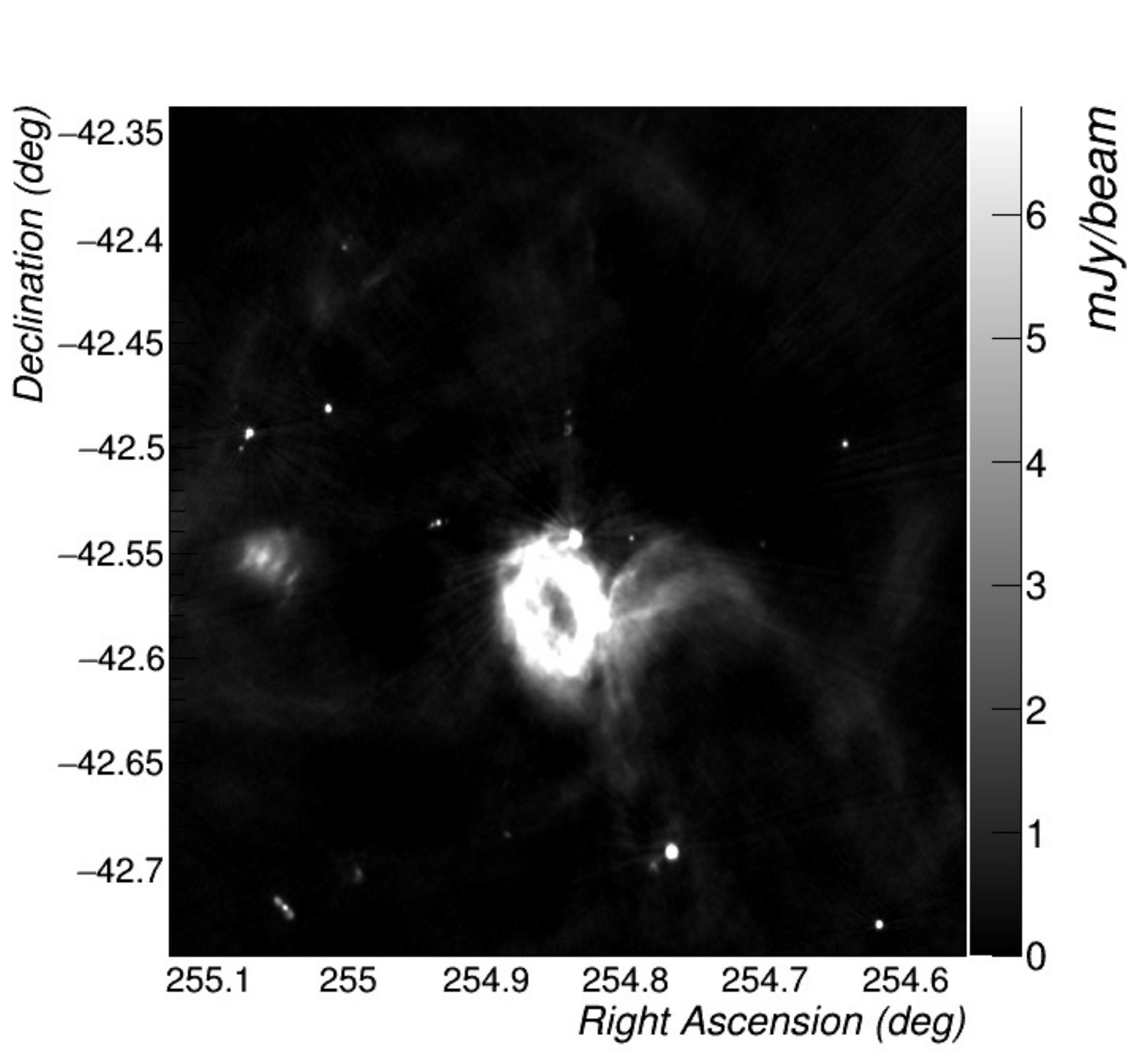}}
\hspace{0.2cm}
\subfloat[Field B\label{ScorpioFieldFig2}]{\includegraphics[scale=0.28]{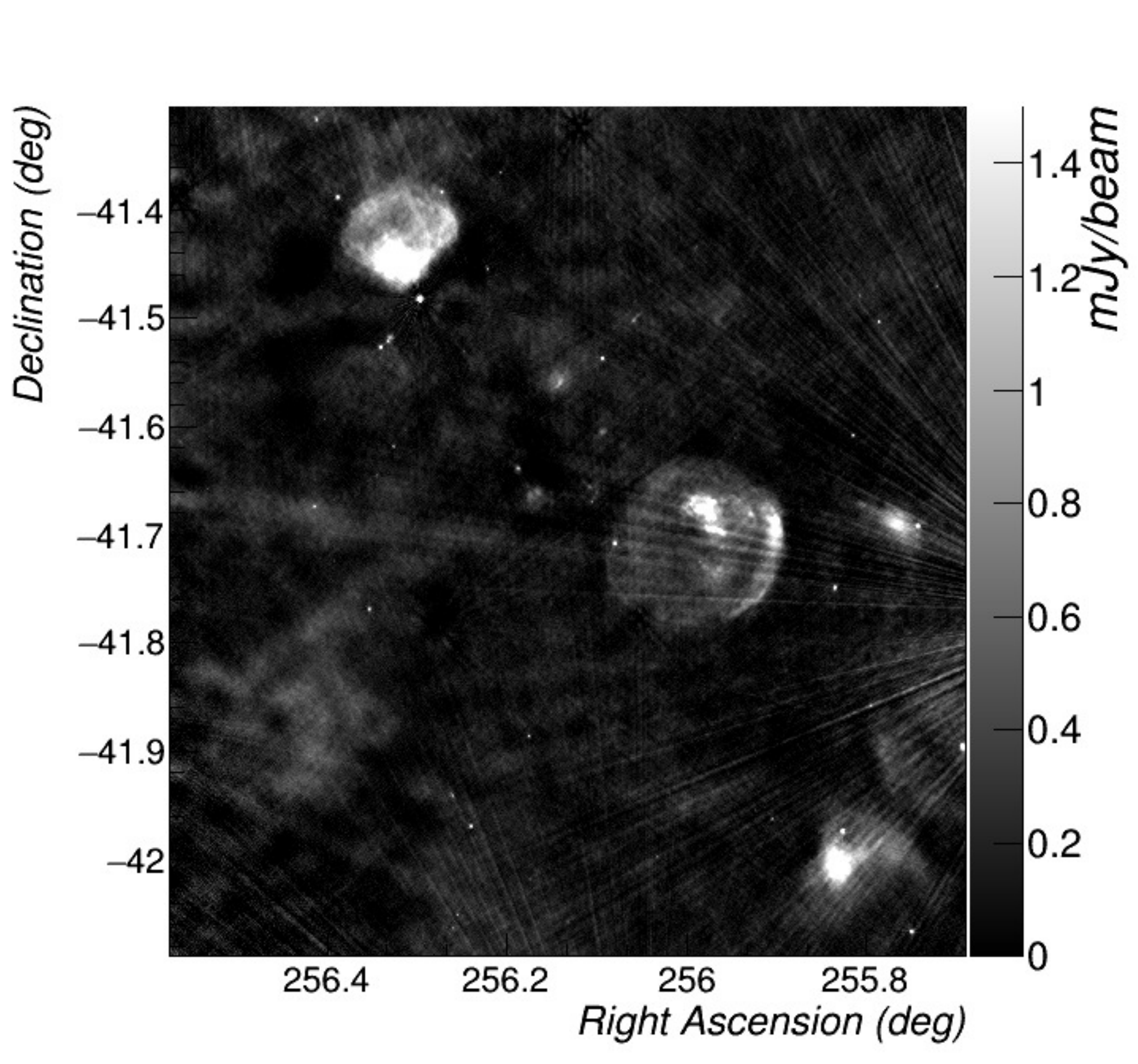}}\\
\vspace{-0.4cm}
\subfloat[Field C\label{ScorpioFieldFig3}]{\includegraphics[scale=0.28]{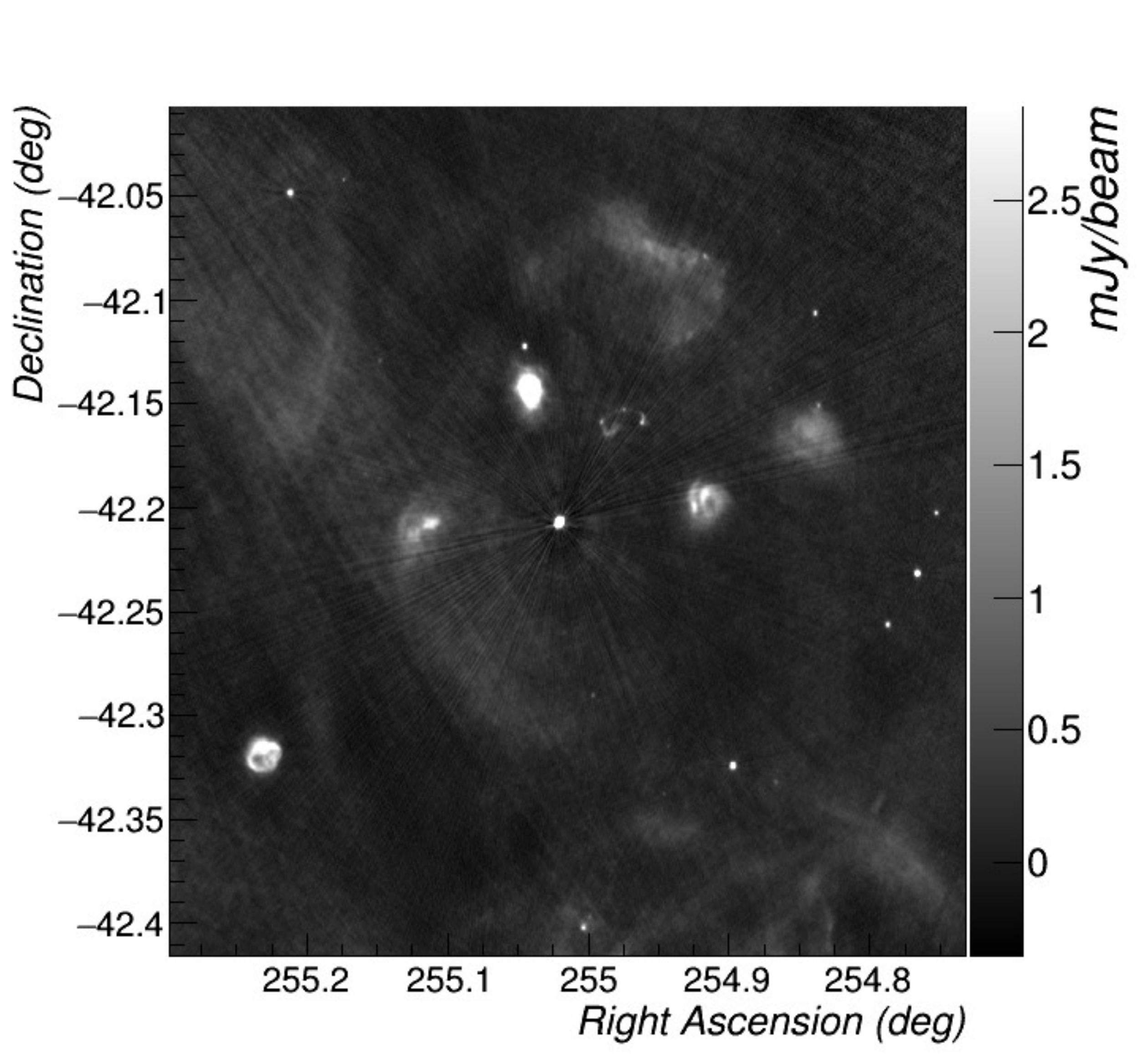}}
\hspace{0.2cm}
\subfloat[Field D\label{ScorpioFieldFig4}]{\includegraphics[scale=0.28]{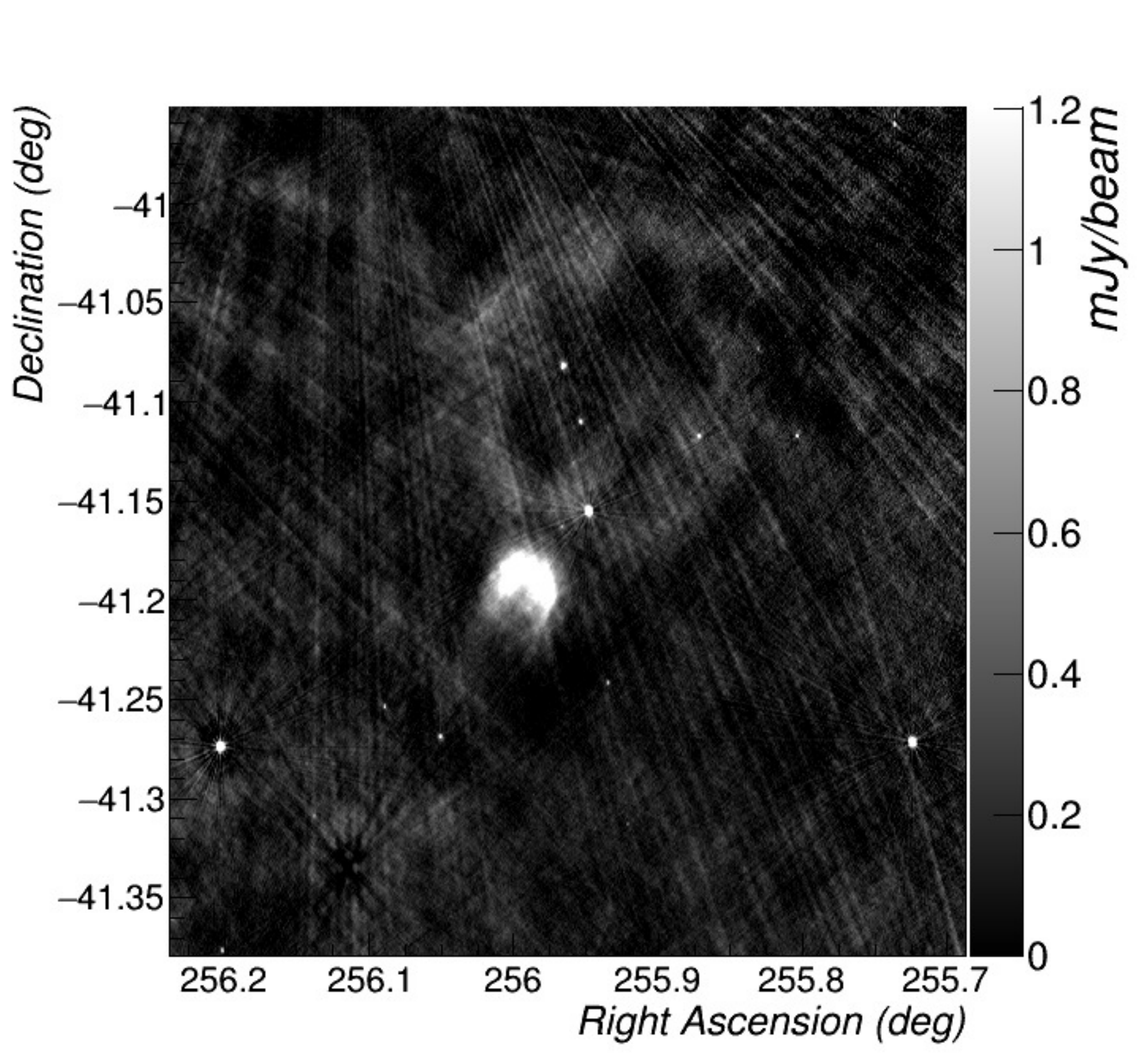}}
\caption{Sample SCORPIO fields (A-D) selected for algorithm testing. Flux units are reported in the z axis.}
\label{ScorpioFieldFig}
\end{figure*}

\subsection{Background and noise estimation}\label{BkgFindingSect}
As noted in Paper I, both background and noise levels are subjected to variations throughout the image, due for example to diffuse emission around 
the Galactic plane or to the accuracy of the image reconstruction. Background and noise information are therefore estimated on a local basis using two alternative methods. 
The first conventional method assumes a rectangular grid of sample pixels and computes the local background and noise levels over a sampling box, centered around each grid center.
Robust background/noise estimators are generally considered to reduce the bias caused by the possible presence of sources falling in the sampling box. 
For instance \textit{Selavy} \citep{Whiting2009,Whiting2012} uses the median and mean absolute deviation from the median (MAD), while the inter-quartile range is adopted in 
\textit{Aegean} \citep{Hancock2012}. Other methods use the previous estimators iteratively clipped up to reach a pre-specified tolerance, as 
in \textit{SExtractor} \citep{Bertin1996} or in Paper I. Several estimators are available in our program: median/MAD, biweight or $\sigma$-clipped estimators. 
Finally, a bicubic interpolation stage is carried out to derive local estimates on a pixel-by-pixel basis, e.g. the background and noise maps.

The second method exploits the pixel spatial information, neglected by the conventional approach, along with the pixel intensity distribution to produce less biased noise/background 
estimates. Two different approaches were implemented. In the first, a superpixel partition of the image is generated (see Section~\ref{SegmentationAlgorithmSection} for more details) 
with region size assumed comparable to the synthesised beam size. An outlier analysis, based on a robust estimate of the Mahalanobis distance \citep{Rousseeuw1990} on region median-MAD parameter space, is then performed 
to detect significative regions (both positive or negative excesses), typically associated with sources or artefacts. Pixels belonging to that regions are marked and 
excluded from the background evaluation. The background and noise maps are finally computed as above by interpolating a robust estimator computed over background-tagged 
pixels in sampling boxes sliding through the entire image. 

A second approach uses a flood-filling algorithm to detect and iteratively clip blobs at some predefined significance level (e.g. 5$\sigma$) with respect to the first 
level estimate of the background and noise maps. Background and noise maps are re-computed at each iteration stage as described above. One or two iterations are typically sufficient.

In practice, the first method can be safely used for bright compact source filtering, in which the 
background estimation is not requested to be highly accurate. The second method should be instead preferred in the search of 
faint compact sources or when thresholding extended bright sources.

The size of the sampling grid is conventionally chosen to achieve sufficient interpolation accuracy at moderate computational cost. Instead, the choice of the box size is often given 
in terms of the beam size (e.g. 10 or 20 larger than the synthesised beam) and may have a considerable impact in the source extraction step: estimates computed on a small 
box could be severely biased by the presence of a source filling the box, while, on the other hand, a too large box could completely smooth out the local 
background/noise variations. In \citet{Huynh2012} the authors compared maps 
obtained by popular source finders, such as \textit{SFind} \citep{Hopkins2002}, \textit{SExtractor} \citep{Bertin1996} and \textit{Selavy} \citep{Whiting2009,Whiting2012}, and 
investigated the optimal parameter settings both for real and simulated data sets. However, they note that a completely automated procedure for background estimation, possibly 
independent on the distribution of sources, is still of crucial importance for future surveys. 

\begin{figure*}
\subfloat[][Field A]{\includegraphics[scale=0.28]{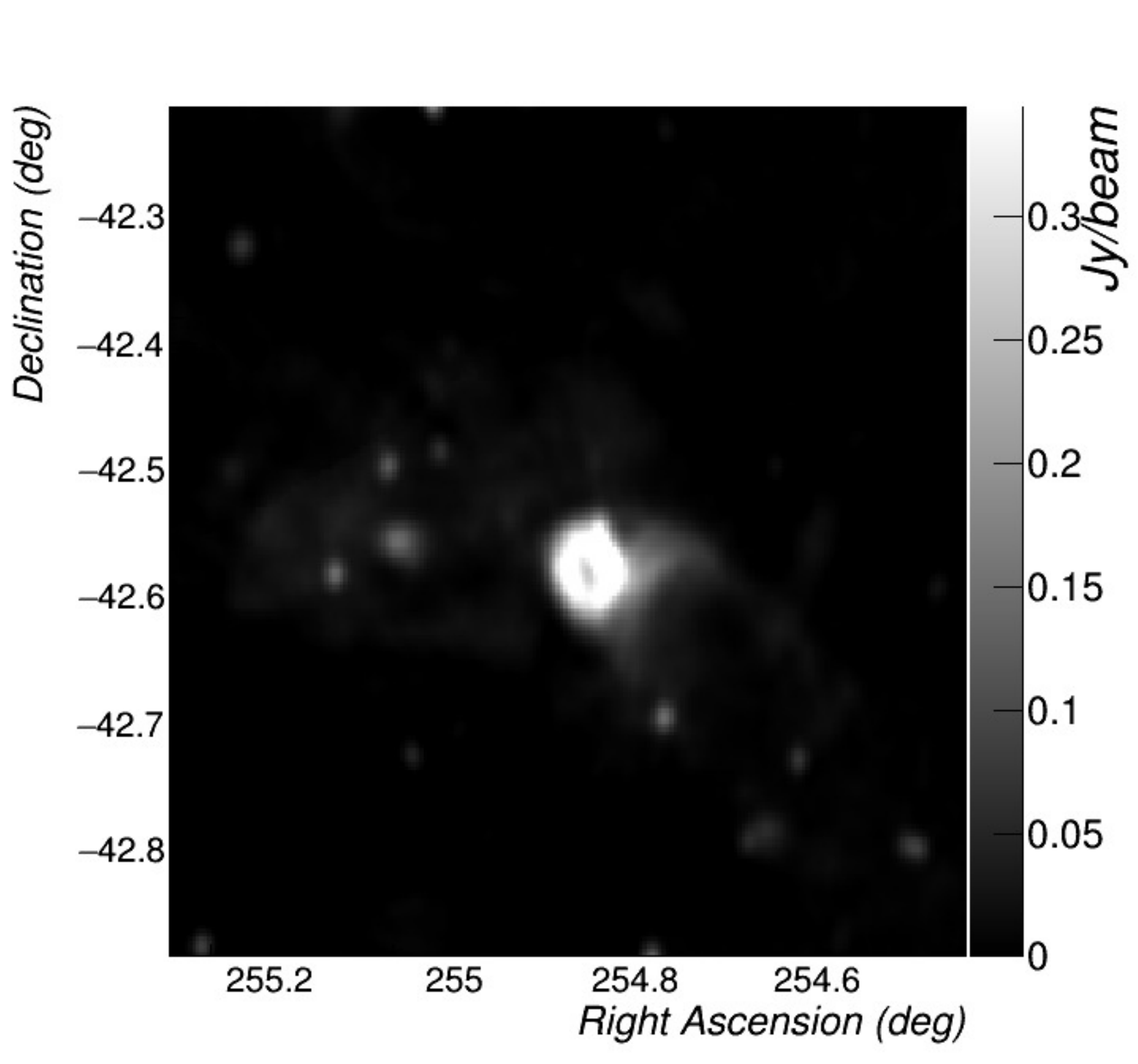}\label{MolongloScorpioFieldFig1}}
\hspace{0.1cm}
\subfloat[][Field B]{\includegraphics[scale=0.28]{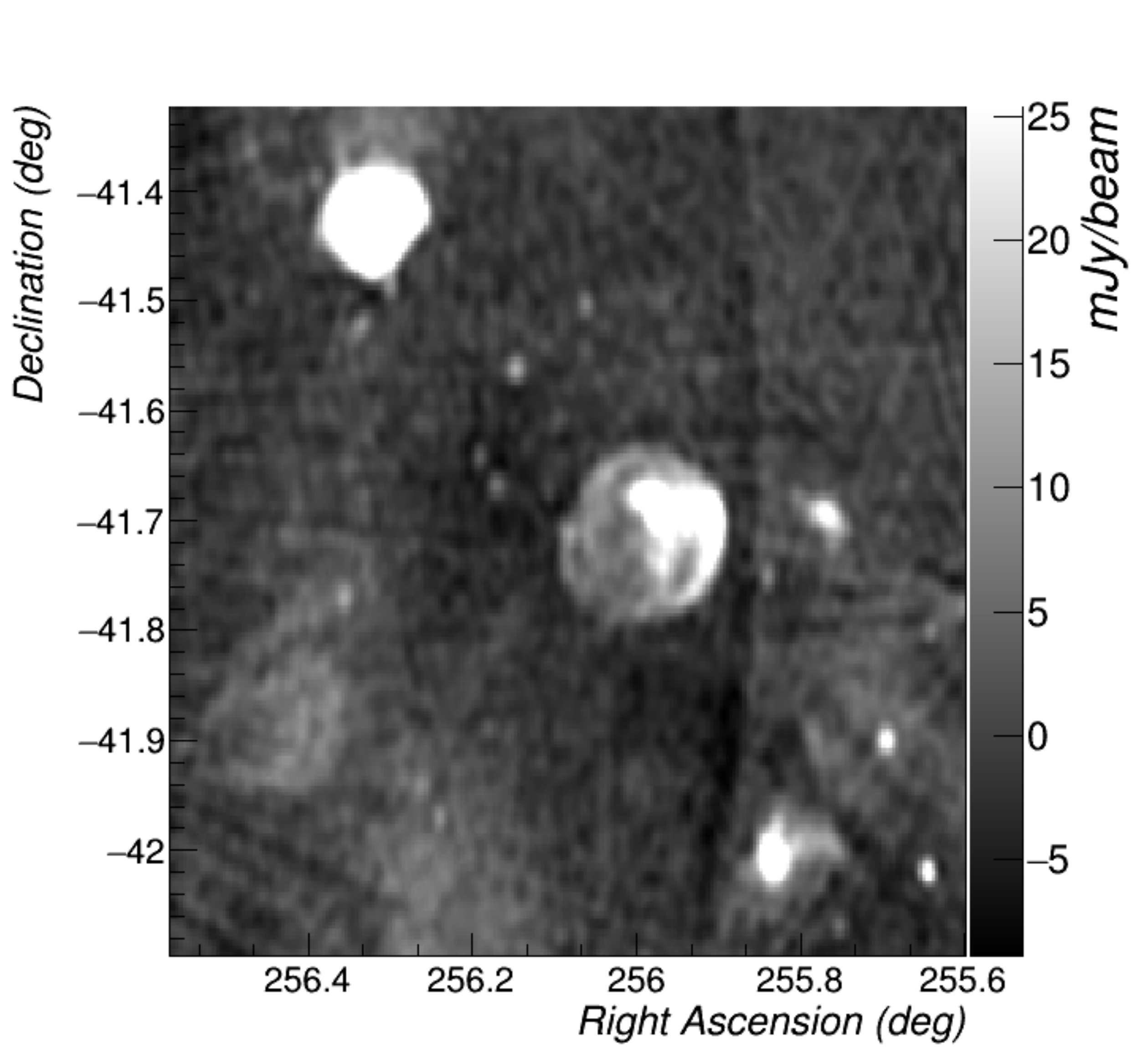}\label{MolongloScorpioFieldFig2}}\\
\vspace{-0.4cm}
\subfloat[][Field C]{\includegraphics[scale=0.28]{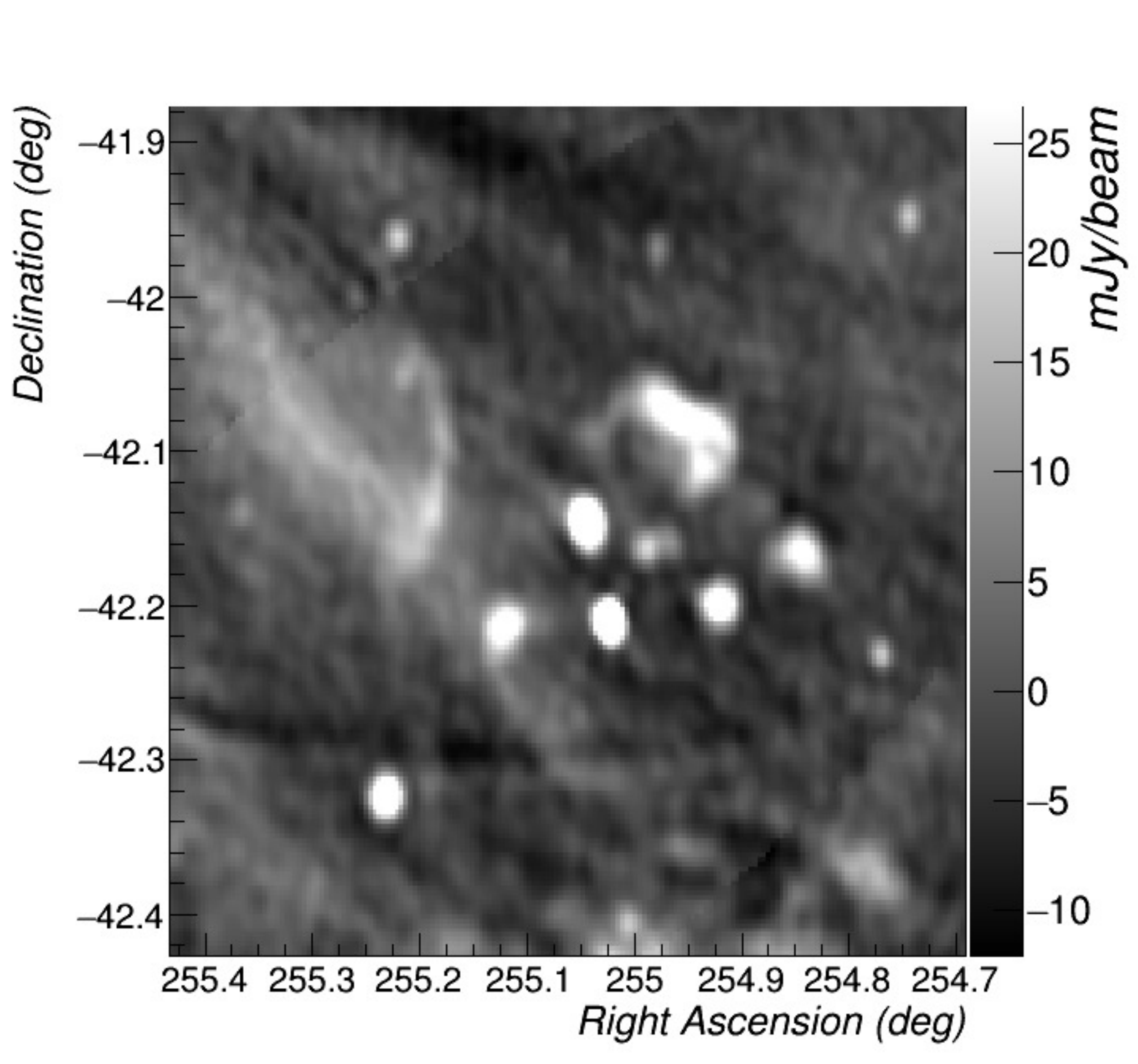}\label{MolongloScorpioFieldFig3}}
\hspace{0.1cm}
\subfloat[][Field D]{\includegraphics[scale=0.28]{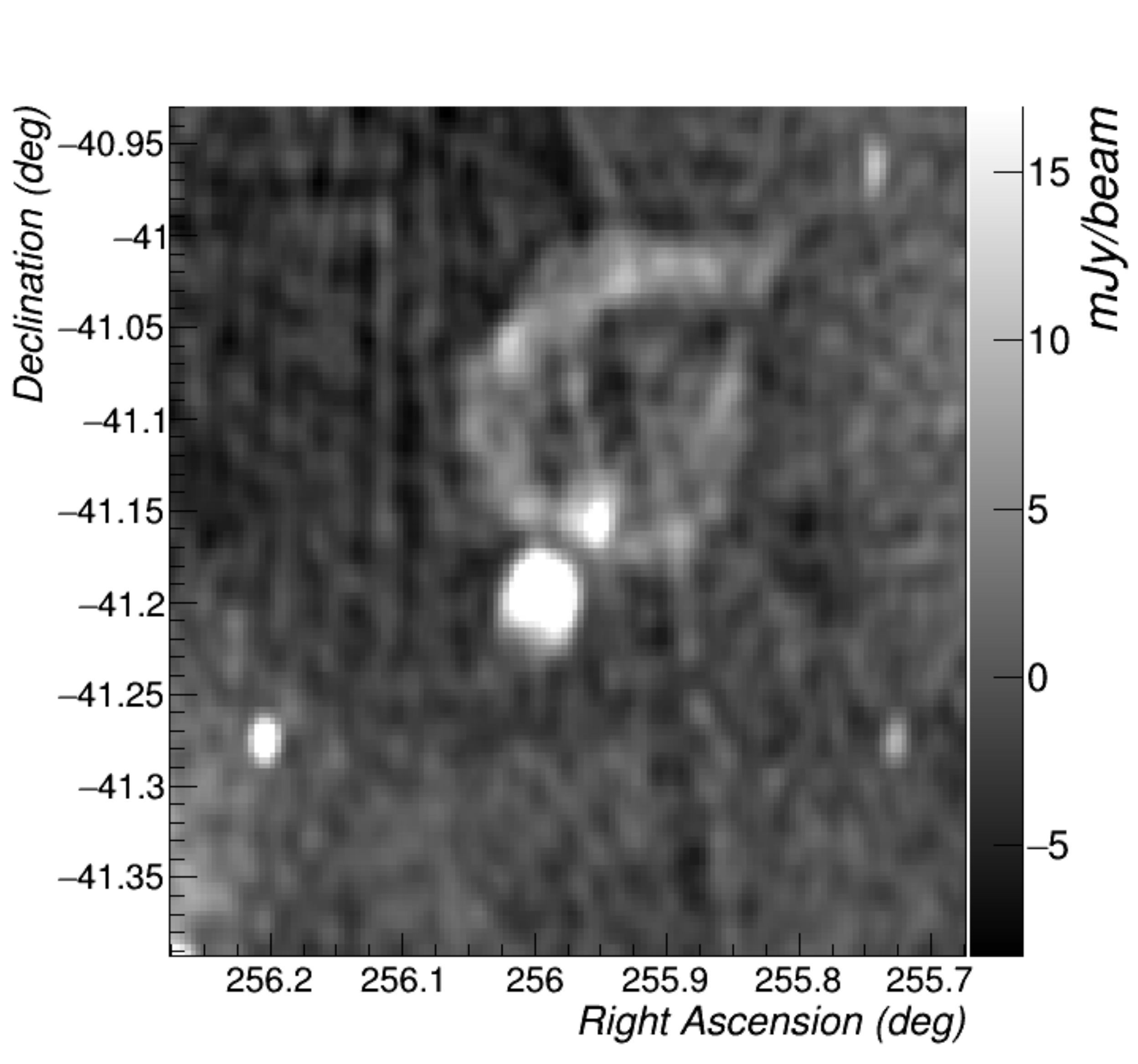}\label{MolongloScorpioFieldFig4}}
\caption{Sample fields (A-D) selected for algorithm testing as observed in the Molonglo Galactic Plane Survey. Flux units are reported in the z axis.}
\label{MolongloScorpioFieldFig}
\end{figure*}

\subsection{Filtering compact sources}\label{CompactSourceFindingSect}
The presence of bright sources in the image significantly hardens the extended source detection task. We therefore implemented a filtering stage to remove them, based on the 
following steps. Blobs of connected pixels are first extracted from the image assuming a flood-filling procedure similar to that carried out in \textit{Aegean} \citep{Hancock2012} and 
\textit{Blobcat} \citep{Hales2012} source finders. A high seed threshold above the computed background is assumed, e.g. 10$\sigma$, and pixels are aggregated down to a merge 
threshold, e.g. 2.6$\sigma$. Each detected blob is subjected to a further search to identify nested blobs. These are extracted by thresholding the image 
curvature map $\kappa$, obtained by convolving the image with a Logarithm-of-gaussian (LoG) kernel, at some pre-specified threshold level (e.g. $\kappa >$0) or adaptively.
A 2-level hierarchy of blobs is finally obtained. 

A set of morphological parameters (e.g. contour parameters, moments, shape descriptors, etc), is computed over the detected blobs and selection cuts are applied to 
identify point-like candidate sources. For example, blobs with a number of pixels that is too large or with an anomalous elongated shape typically fail to pass the point-like cut.

Blobs tagged as "point-like" are removed from the input image using a morphological dilation operator with configurable kernel shape (e.g. elliptic or squared) and size, as suggested 
in \citet{Peracaula2015}, and replaced with a random background realization. A kernel size larger than 5 pixels was assumed to prevent the source halo pixels to further 
affect the residual image.

\begin{figure*}
\subfloat[Field A]{\includegraphics[scale=0.22]{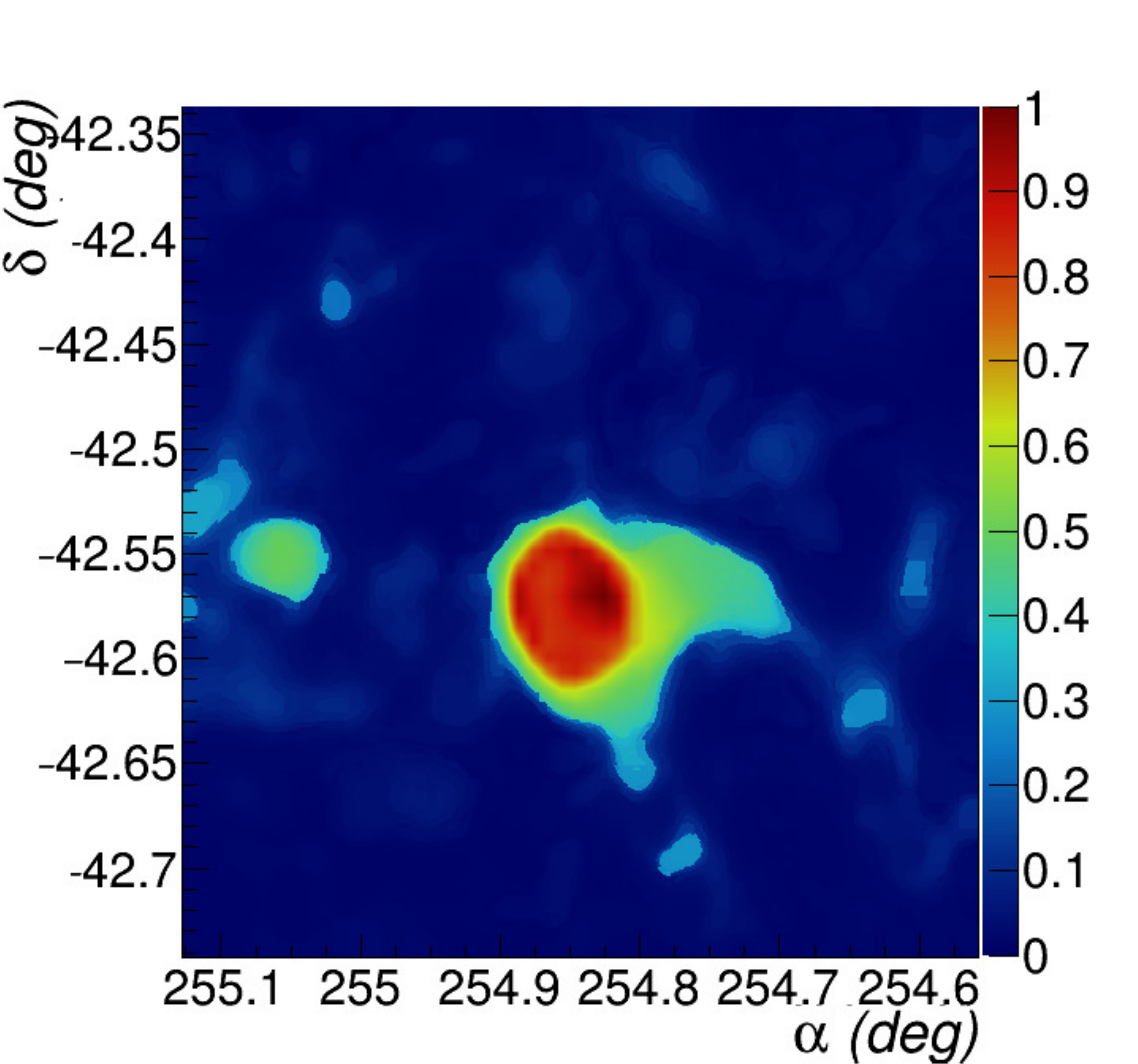}}
\hspace{-0.cm}
\subfloat[Field A]{\includegraphics[scale=0.22]{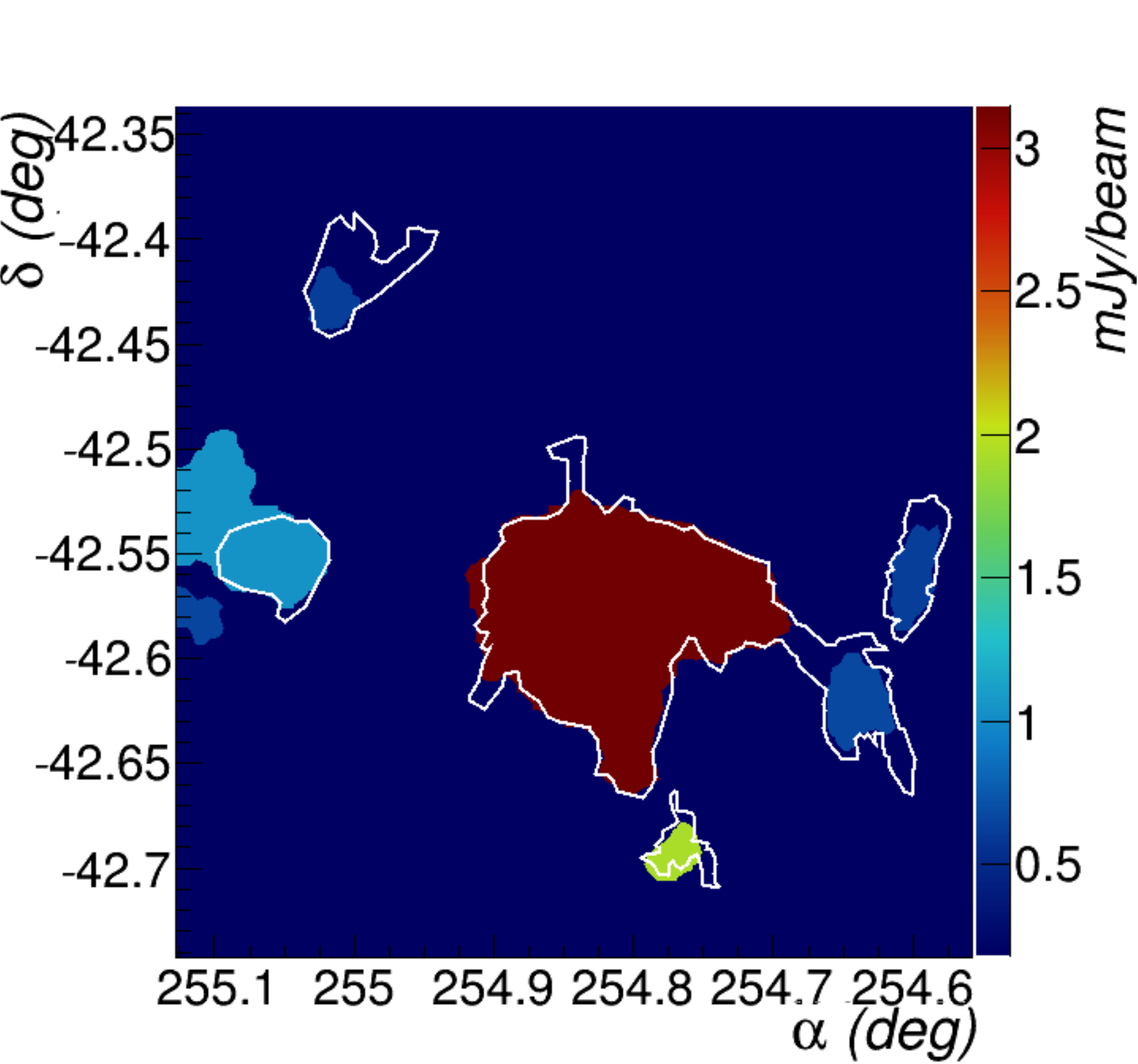}}\\
\vspace{-0.4cm}
\subfloat[Field B]{\includegraphics[scale=0.22]{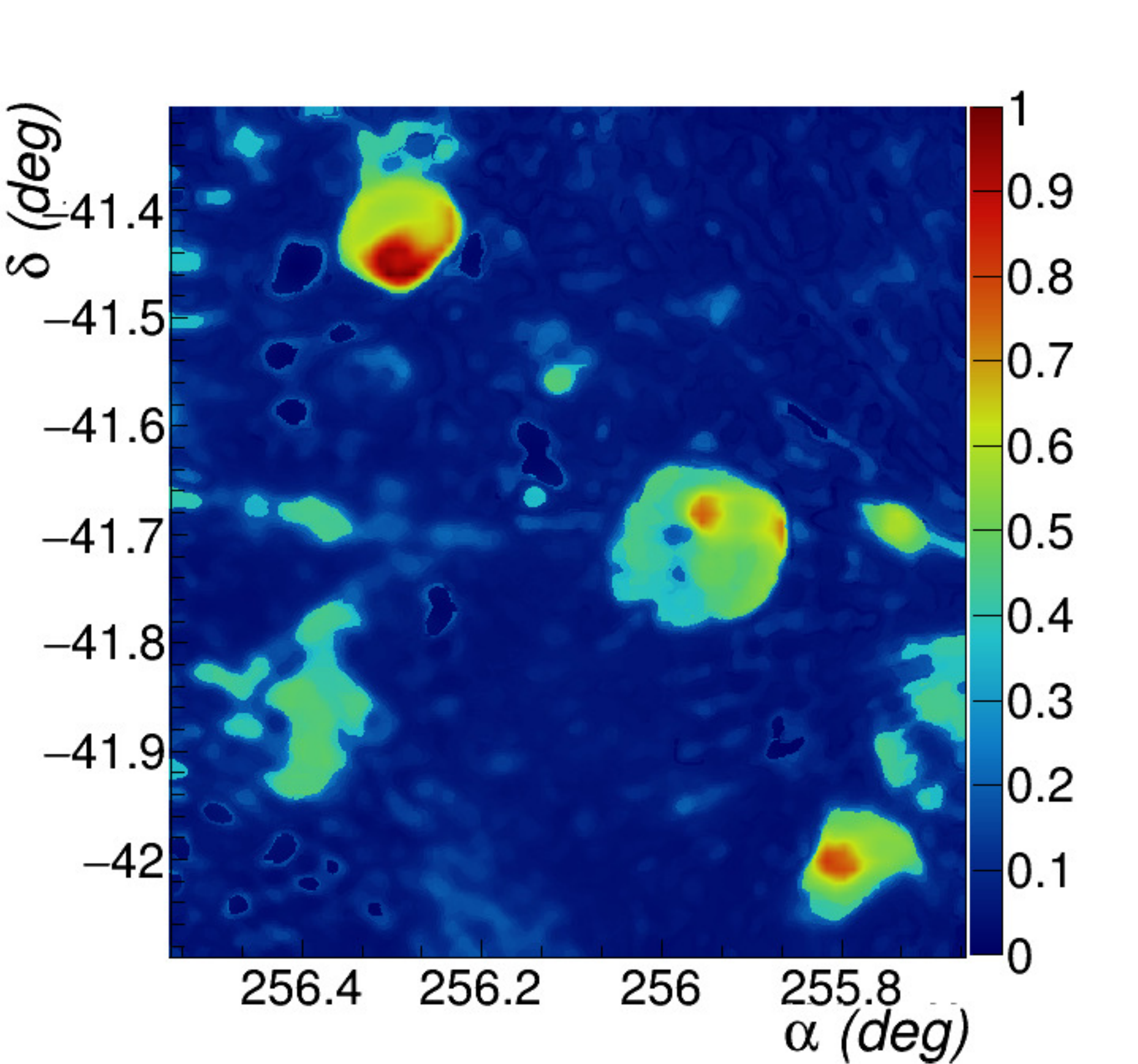}}
\hspace{-0.cm}
\subfloat[Field B]{\includegraphics[scale=0.22]{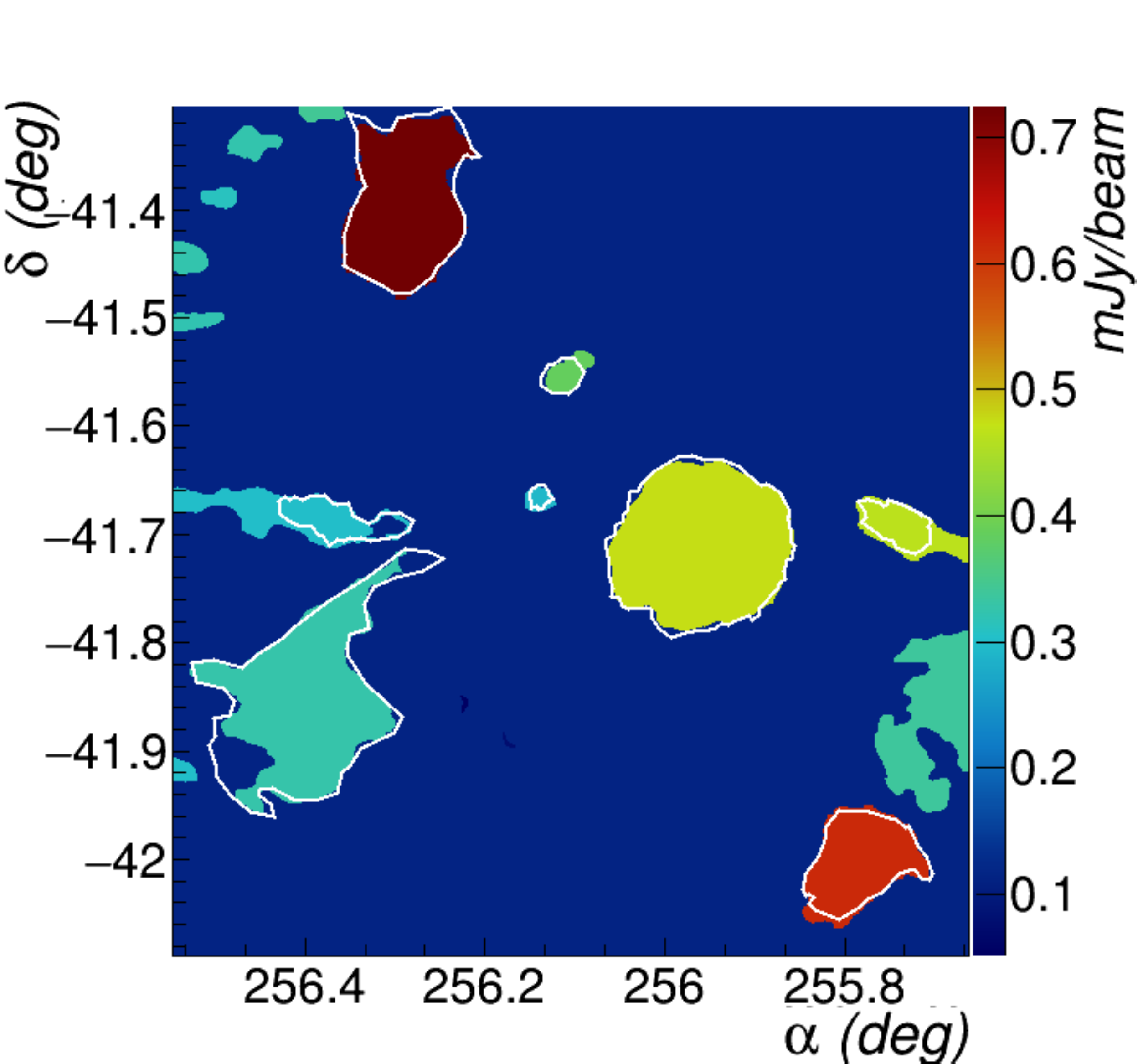}}\\
\vspace{-0.4cm}
\subfloat[Field C]{\includegraphics[scale=0.22]{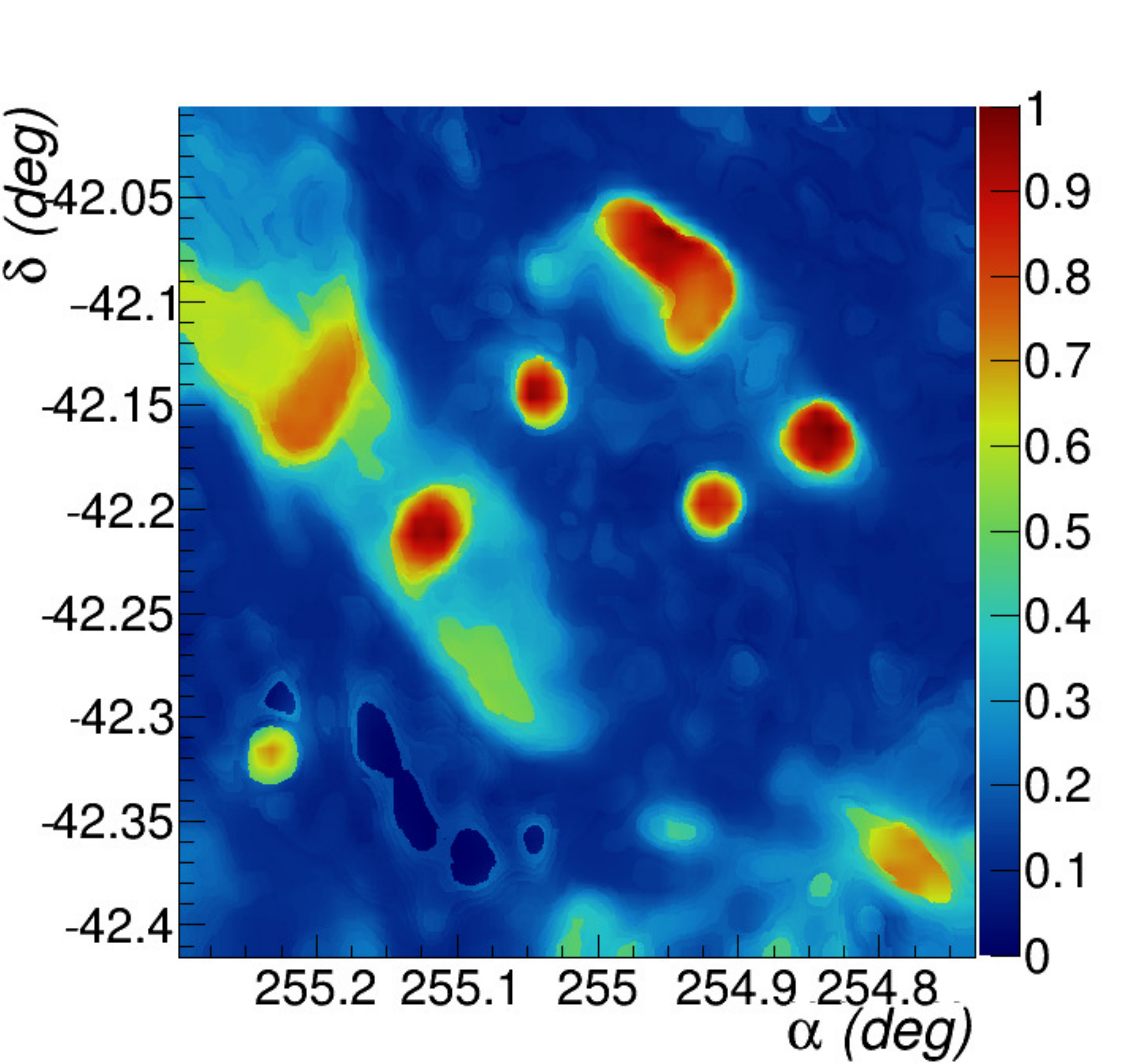}}
\hspace{-0.cm}
\subfloat[Field C]{\includegraphics[scale=0.22]{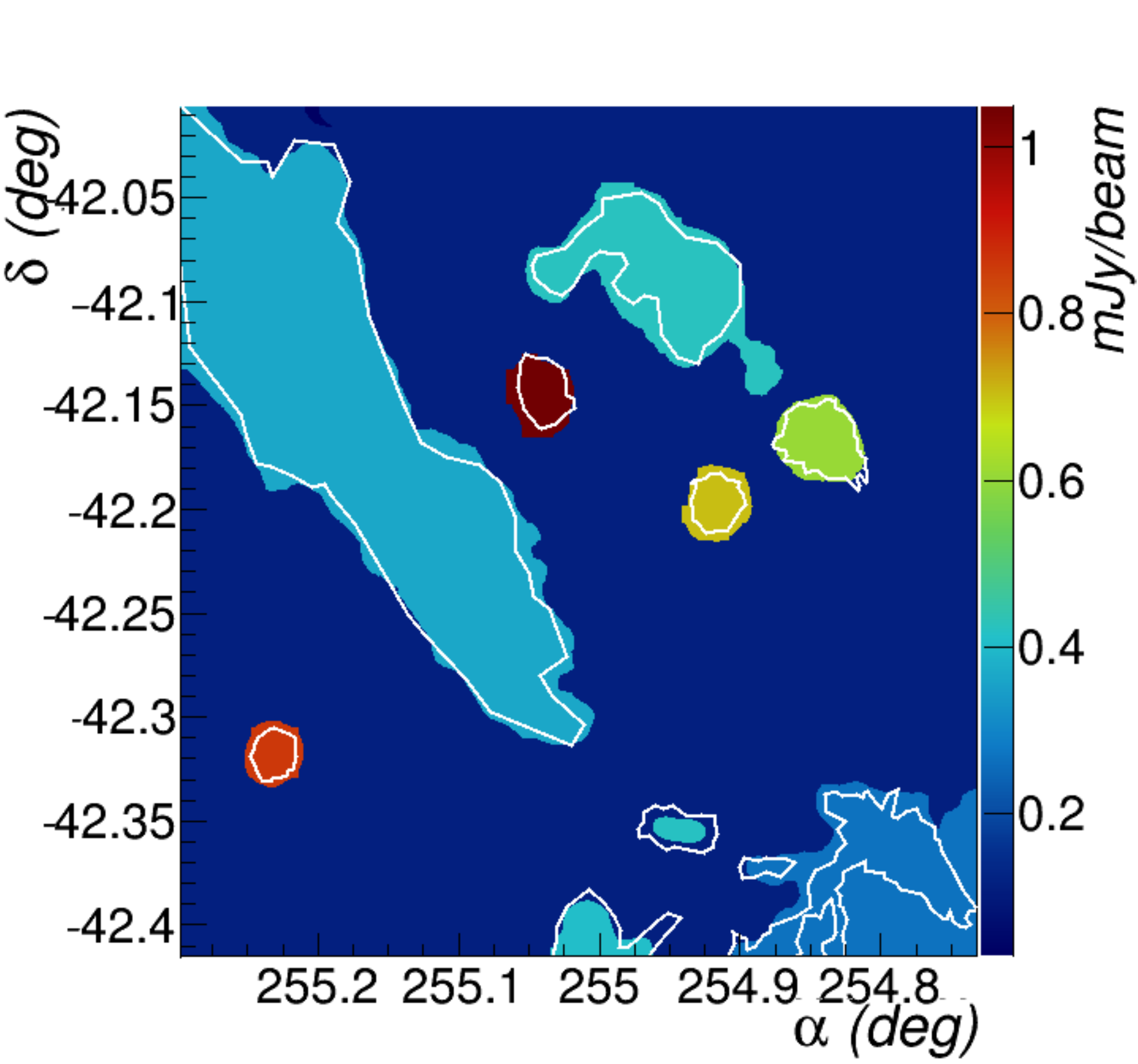}}\\
\vspace{-0.4cm}
\subfloat[Field D]{\includegraphics[scale=0.22]{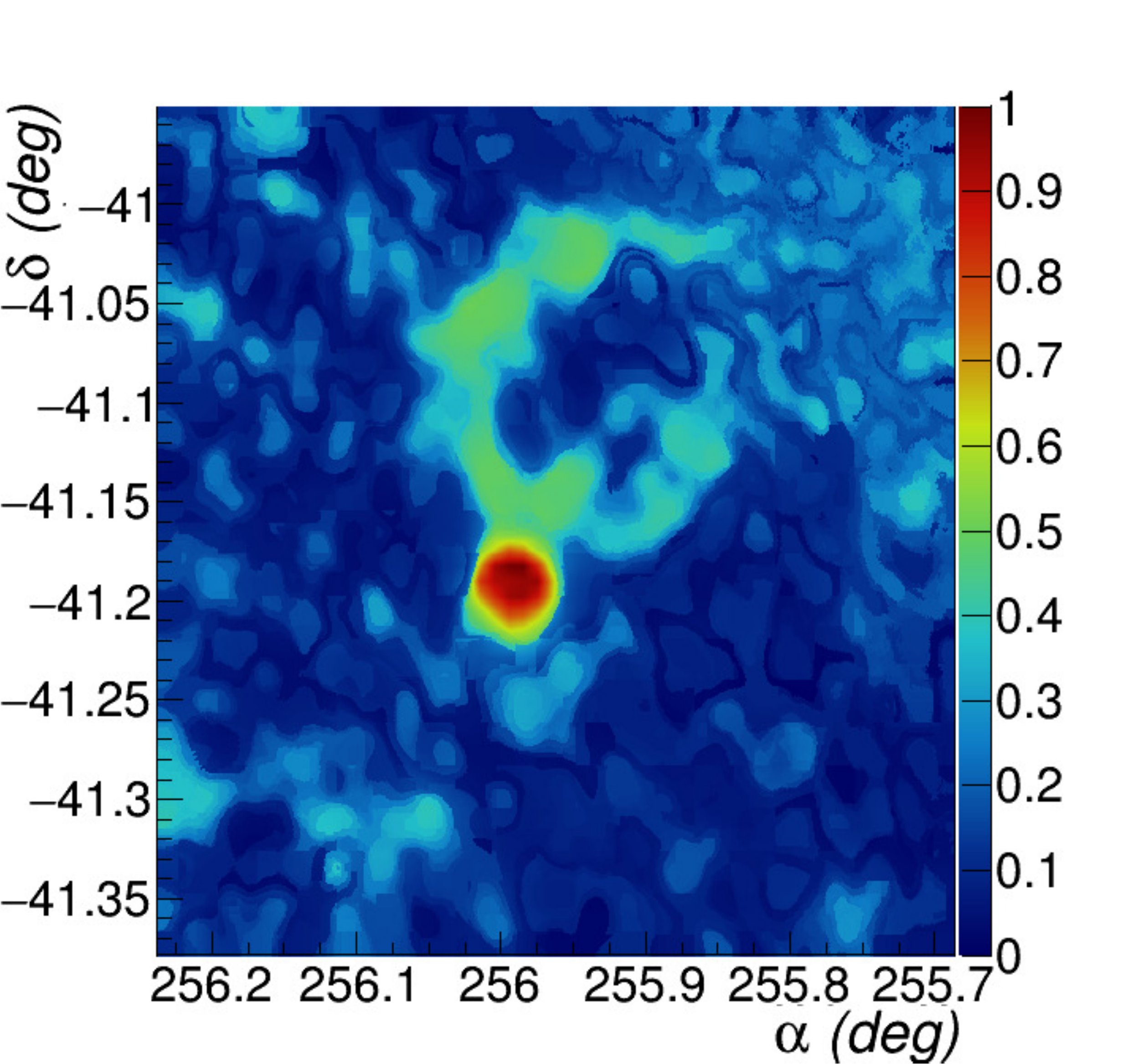}}
\hspace{-0.cm}
\subfloat[Field D]{\includegraphics[scale=0.22]{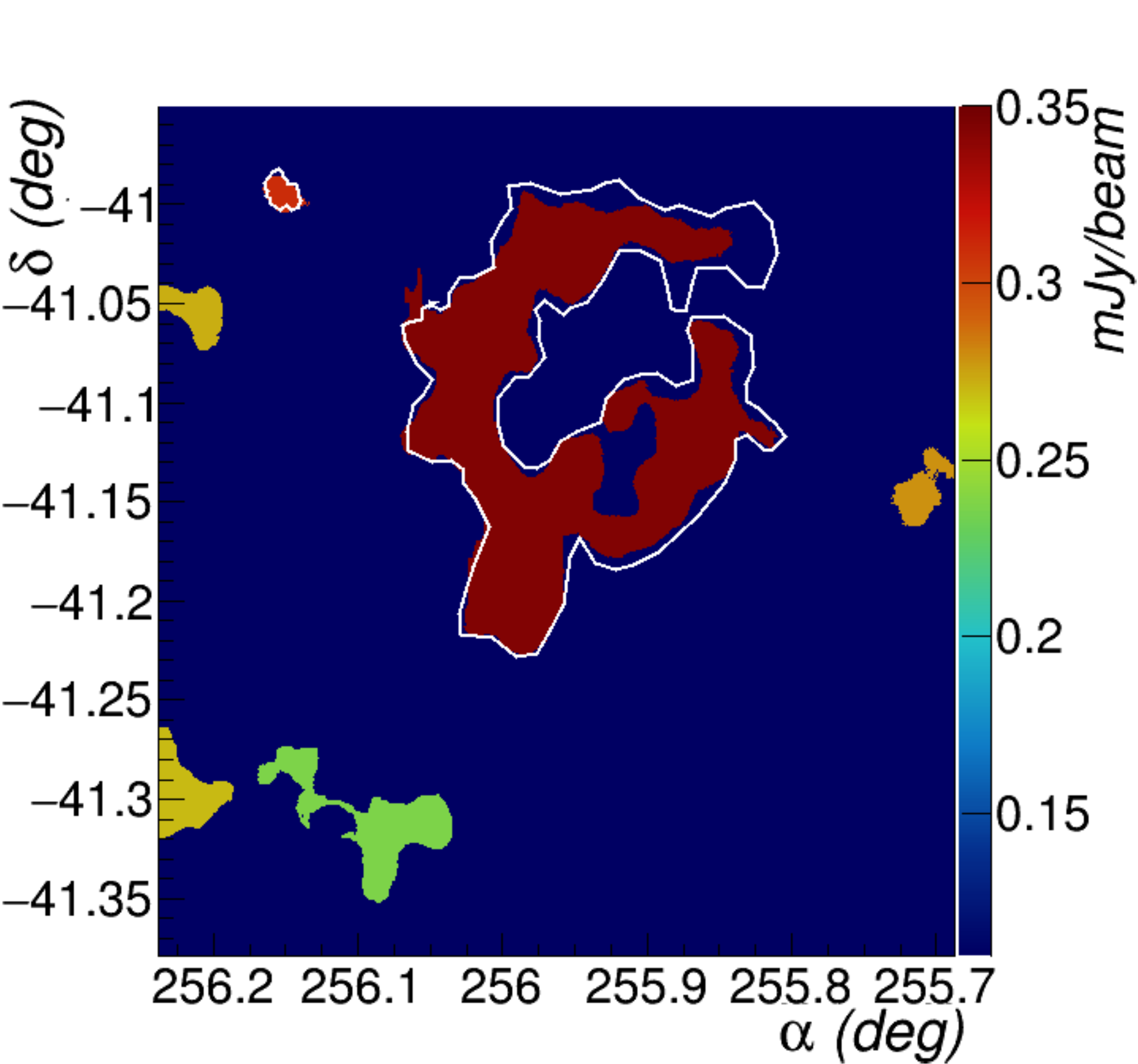}}
\caption{Segmentation results obtained for the test fields A-D (from top to bottom) assuming $l$=20 and $\beta$=1 (see text). 
Left: Saliency maps normalized to range [0,1]; Right: Segmentation maps. Each segmented region is colored in the plot according to the mean of its pixel fluxes in mJy/beam units. 
The white contour lines correspond to a manual segmentation generated by an expert astronomer.}
\label{ScorpioSegmResultsFig}
\end{figure*}

\subsection{Segmentation algorithm}\label{SegmentationAlgorithmSection}
We developed a segmentation algorithm for extraction of extended sources, based on a superpixel segmentation algorithm followed by a 
hierarchical clustering stage to aggregate similar segments into final candidate source regions. The algorithm steps are described below and a summary of the relevant algorithm 
parameters is reported in Table~\ref{AlgorithmParTable}:
\begin{enumerate}
 \item \emph{Initialization}: Compute a set of filtered images to be used during the clustering stage, namely the image curvature $\kappa$ and an edge-sensitive map $\psi$. The latter 
 can be alternatively obtained by convoluting the input image with a set of Kirsch filters oriented along different directions or as the result of the Chan-Vese contour finding algorithm 
 \citep{ChanVese2001}.

 \begin{figure*}
\subfloat[Field E]{\includegraphics[scale=0.28]{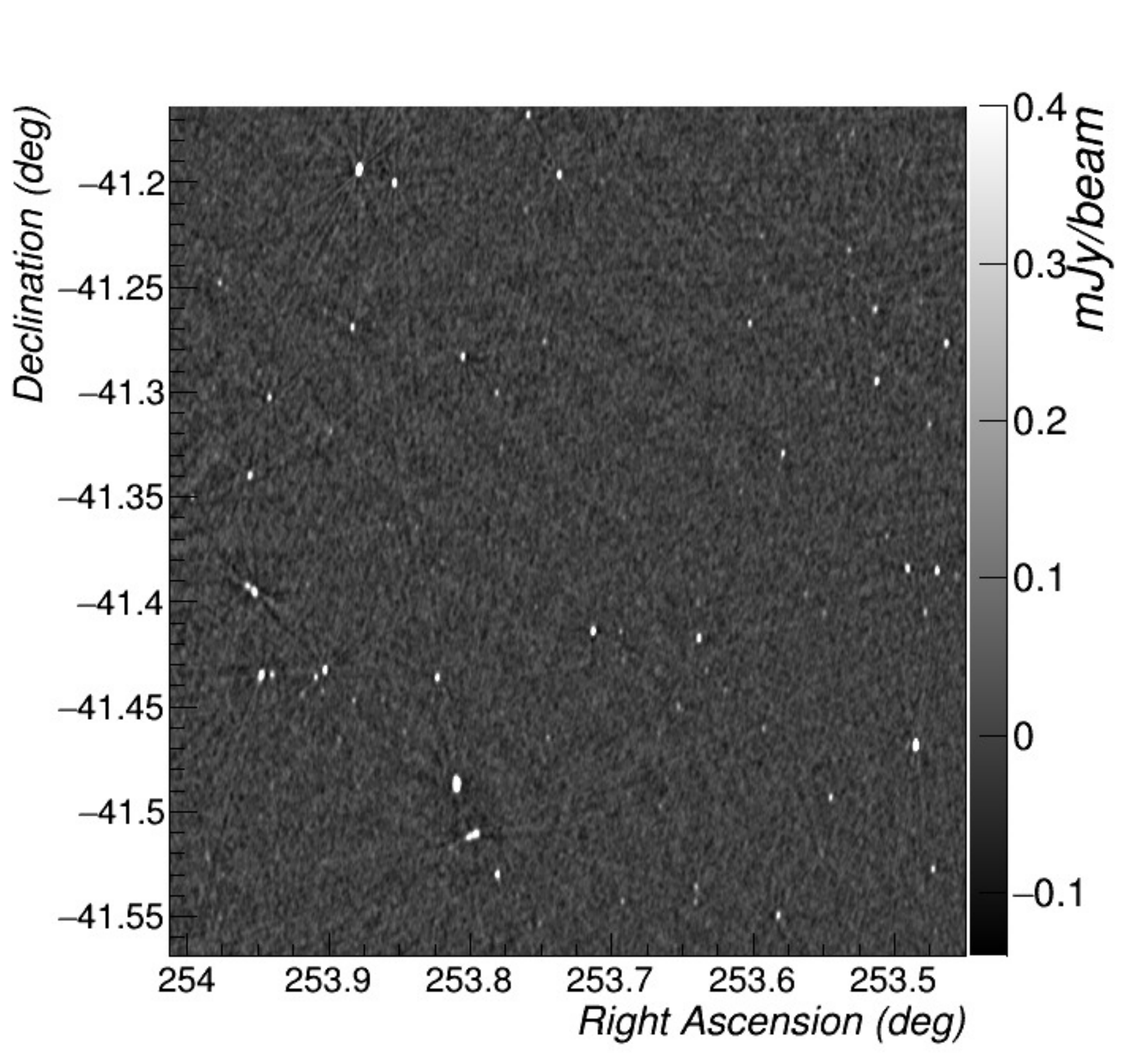}}
\hspace{-0.cm}
\subfloat[Field E - Residual]{\includegraphics[scale=0.28]{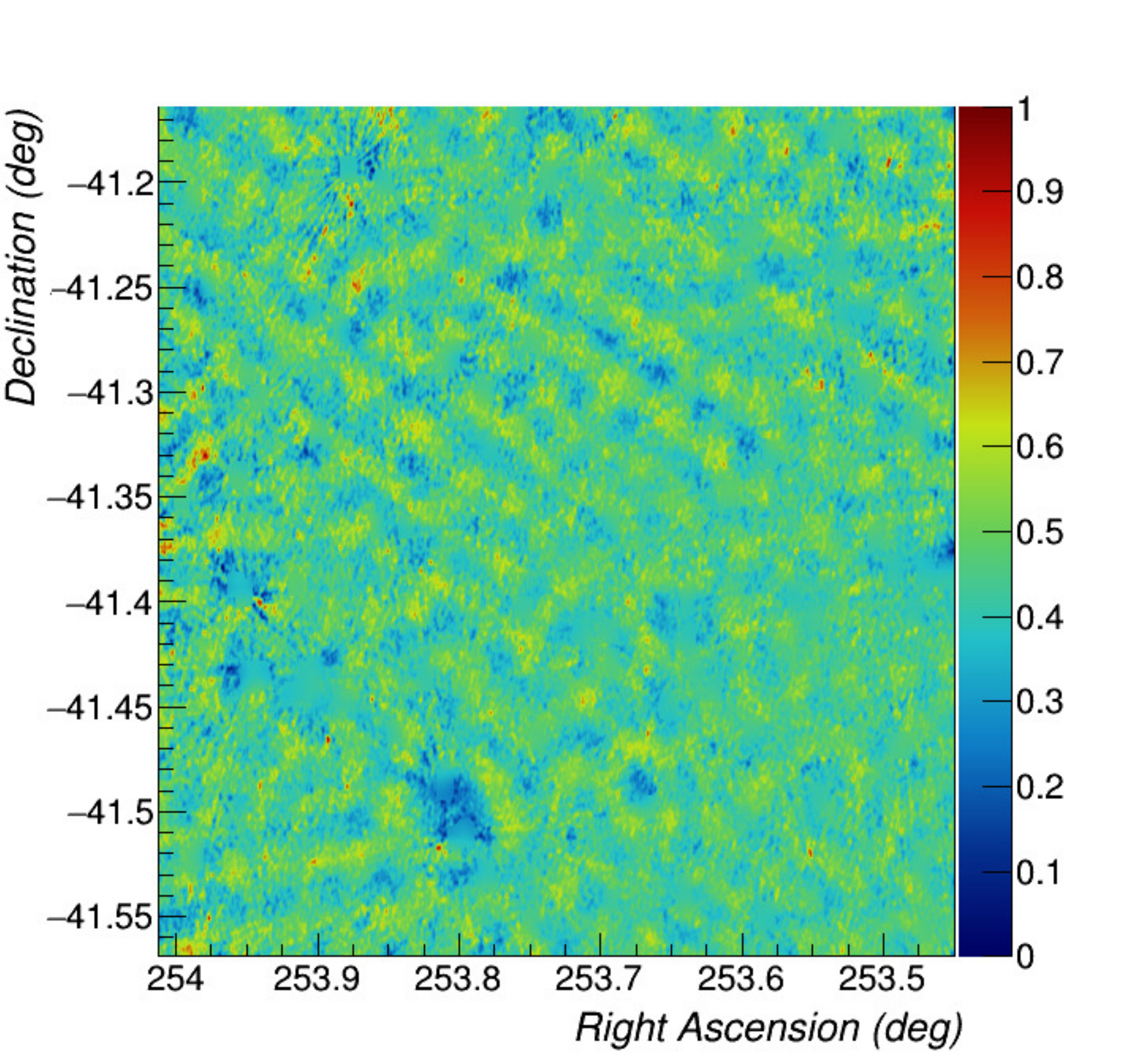}}
\hspace{-0.cm}
\caption{Left: Sample SCORPIO field E selected for algorithm testing. Flux units are reported in the z axis; Right: Residual map, normalized to range [0,1], obtained after  
applying point-like source and smoothing filtering stages to the input map.}
\label{SparseFieldResultsFig}
\end{figure*}

 \item \emph{Superpixel segmentation}: In this stage the image is over-segmented into $N_{\texttt{R}}$ connected regions or superpixels using flux and spatial information as input 
 observables. To this aim we made use of the \textit{Simple linear iterative clustering (SLIC)} algorithm developed by \citet{Achanta2012}, which uses the k-mean algorithm to 
 cluster pixels according to an intensity and spatial proximity measure. Segmentation is controlled by a set of input parameters, such as the desired 
 superpixel size $l$, typically fixed to the smallest detail to be distinguished (e.g. close to the beam size to detect compact sources or larger to search for 
 extended sources), the minimum number of pixels in a region ($N_{\texttt{min}}$) and a regularization parameter $\beta$ balancing spatial and intensity clustering in the 
 distance measure $D_{ij}$ between a pixel $i$ and a superpixel center $j$: 
 \begin{equation}
 D_{ij} = \sqrt{ D_{ij,c}^2 + \left(\frac{\beta}{l\times l}\right)^2 D_{ij,s}^2}
 \end{equation}
 $D_{ij,c}$ and $D_{ij,s}$ being the intensity and spatial Euclidean distances between pixel $i$ and superpixel $j$. Higher $\beta$ enhances the spatial proximity and favors more compact 
 superpixels in the initial partition. In turn, lower $\beta$ favors clustering in intensity and superpixels with less regular shapes but adhering more tightly 
 to the object contours.
 
 For each region $i$ an appearance parameter vector $\mathbf{x}_{i}=(\mu_{i},\sigma_{i},\mu_{i,\kappa},\sigma_{i,\kappa})$ 
 is computed, with $\mu_{i}$ and $\mu_{i,\kappa}$ denoting 
 respectively the mean of flux and curvature of pixels belonging to region $i$, while $\sigma_{i}$ and $\sigma_{i,\kappa}$ are their standard deviations. 
 With this parameter choice, the computation and update of the region parameters after a merging can be done iteratively in a very fast way, namely without partially sorting 
 the region pixel vector as in the case of median and MAD estimators. 
 
 \item \emph{Saliency map estimation}: A saliency map is estimated in this step to enhance significant objects in the input image with respect to the background. Following 
 \citet{Zhang2013}, a saliency estimator $S_i$ is computed for each region as:
 \begin{equation}
 S_i= 1-\exp\left(-\frac{1}{K}\sum_{j=1}^{K}\delta_{ij}\right)\;\;\;\;\;\;\;\;\;\;\;\delta_{ij}=\frac{d_{ij,c}}{1+d_{ij,s}}
 \end{equation}
 where $d_{ij,c}$ is the Euclidean distance between appearance vectors $\mathbf{x}_{i}$ and $\mathbf{x}_{j}$ of region $i$ and $j$, $d_{ij,s}$ the distance between their centroids. The sum is 
 computed over the $K$ nearest neighbors of region $i$, typically 10\% or 20\% out of the total number of regions. Salient objects are likely to have similar pixels more confined 
 in space compared to similar pixels belonging to the background which 
 are more spatially spread in the image. To detect salient features at different scales, we combined saliency maps computed at different resolutions, e.g. corresponding to 
 initial partitions with different superpixel sizes. Finally, multi-resolution saliency maps are combined with the computed local noise and background maps, which 
 are found to be also sensitive to the diffuse emission. A saliency map with almost full pixel resolution is finally determined.

 \item \emph{Superpixel tagging}: Each pixel $i$ is tagged as background/object/untagged candidate if its saliency $S_i$ is within some adaptive threshold levels:
 \begin{equation}
    S_i= \left\{ 
      \begin{array}{ll}
	\texttt{background} & S_i<S_{\texttt{thr}}^{\texttt{bkg}}  
	\vspace{0.2cm}
	\\  
	\texttt{object} & S_i>S_{\texttt{thr}}^{\texttt{sig}}
	\vspace{0.2cm}
        \\ 
	\texttt{untagged} & \text{otherwise}
      \end{array} 
    \right.
\end{equation}
Different saliency thresholding approaches are possible. One of the most used in saliency studies  \citep{Achanta2009,Perazzi2012,Kim2014,Zhang2013} assumes a global adaptive 
threshold of the kind $S_{\texttt{thr}}^{\texttt{bkg,sig}}=f_{\texttt{thr}}^{\texttt{bkg,sig}}\times\langle S\rangle$, where $\langle S\rangle$ is the average (or median) saliency of 
the map and $f$ is a numerical factor (e.g. $f$=1 for the background and $f$=2 for the signal; \citet{Achanta2009,Zhang2013}).
After several tests performed on different maps we obtained optimal results by combining different global threshold measures:
\begin{equation}
S_{\texttt{thr}}^{\texttt{sig}}= \texttt{max}\{f_{\texttt{thr}}^{\texttt{sig}}\times\langle S\rangle,\texttt{min}\{S_{\texttt{thr}}^{\texttt{Otsu}},S_{\texttt{thr}}^{\texttt{valley}} \}\}
\end{equation}
where $S_{\texttt{thr}}^{\texttt{Otsu}}$ is the threshold level computed through the Otsu method (e.g. see \citealt{Sezgin2004} for a review of thresholding methods) 
and $S_{\texttt{thr}}^{\texttt{valley}}$ is the threshold corresponding to the first valley detected in the pixel saliency histogram. 
The threshold level factor $f_{\texttt{thr}}^{\texttt{sig}}$ is chosen as a trade-off between false detection rate and object detection efficiency. 
The alternative approach, more computationally expensive, is employing the local adaptive thresholding method used also for compact source extraction with 
or without outlier rejection. 

Superpixels are finally tagged as background, object or untagged candidates according to the majority of their pixel tags.

\item \emph{Superpixel graph}: Identify 1st- and 2nd-order neighbors to each region $i$=1,\dots,$N_{\texttt{R}}$ and build a corresponding link graph as described in \citet{Bonev2014}. 
By 1st-order neighbors, we denote the regions surrounding and sharing a border with region $i$. For each region link $i-j$ in the graph, compute an edgeness $E_{ij}$ parameter 
related to the amount of edge present on the shared border between region $i$ and $j$. For 1st-order neighbors, this is estimated by taking the average of $\psi$ over the pixels 
located on the shared boundary, while for 2nd-order neighbors, it assumes the largest value present in the $\psi$ map.

Let us consider an asymmetric dissimilarity measure $\Delta_{ij}$ between neighbor regions $i$ and $j$ given by:
\begin{equation}
\Delta_{ij}= (1-\lambda)d(\mathbf{x}_{i},\mathbf{x}_{i\cup j}) + \lambda E_{ij}
\end{equation}
where $d(\cdot,\cdot)$ is the Euclidean distance between feature vectors, $E_{ij}$ the edgeness parameter and $\lambda$ a regularization parameters balancing distance and edgeness weights in 
$\Delta_{ij}$. 

The above measure expresses the change of feature vector $\mathbf{x}_{i}$ caused by a potential merging with region $j$, which is favored when the distance between feature 
vectors is small and penalized when there is a border in between the two regions. Note that $\Delta_{ij}\neq\Delta_{ji}$.

Compute the adjacency matrix $\mathbf{A}$ of the graph with elements $a_{ij}$:
\begin{equation}
a_{ij}= \frac{\Delta_{ij}^{-1}}{\sum_{j}\Delta_{ij}^{-1}}
\end{equation}
properly normalized to express a transition probability from node $i$ to $j$.
  
\item \emph{Superpixel merging}: Following \citet{Ning2010} and \citet{Zhang2013}, merge superpixels on the basis of a maximum similarity criterion by iterating the following steps 
until no more merging is possible:
\begin{enumerate}
\item Merge untagged regions to candidate background regions if their similarity is maximal among neighbor similarities.  
\item Adaptively merge untagged regions if their similarity is maximal among neighbors similarities.
\end{enumerate}
Untagged regions shrink during the previous stage, while background regions grow. Signal-tagged regions are not affected in the previous stages. Superpixel parameter vector and graph 
(neighbor links, dissimilarity/adjacency matrix) are updated after each iterated merging stage. When no more merging is favored, all the remaining untagged regions are labeled as 
signal candidates. This stage always converges to assign all regions to either background or signal. 

A suitable superpixel merging order for each of the steps described above is determined as in \citet{Bonev2014} using the Google PageRank algorithm \citep{Brin1998} on the 
transition matrix $A$, that is solving the following equation:
\begin{equation}
\mathbf{p} = (1-d)\mathbf{e} + d\mathbf{A}^{T}\mathbf{p} 
\end{equation}
in which $\mathbf{p}=(p_{1},p_{2},\dots,p_{N_{\texttt{R}}})$ is the desired vector with rank values (the principal eigenvector of $\mathbf{A}$), $d$ is the damping factor which 
can be set to a value between 0 and 1 (e.g. $d$=0.85 as in \citet{Brin1998,Page1999}) and $\mathbf{e}$ is a column vector of all 1's. The equation is solved 
by using the power iteration method \citep{Golub1983}. $\mathbf{p}$ is sorted and allows to select nodes with higher ranks for merging.

\item \emph{Source selection}: In this step sources are identified from the collection of signal candidate regions selected in the previous stage. 
Following \citet{Bonev2014} the most similar signal regions are hierarchically clustered if their mutual dissimilarities ($\Delta_{ij}$, $\Delta_{ji}$) 
are within a pre-specified tolerance. Only a percentage (e.g. 30\%) of top ranked merging are allowed at each clustering iteration. 

A practical criterion for the merging is allowing first neighbors to always merge (e.g. a sort of flood-fill approach over superpixels) and assuming a tolerance for 
2nd-order neighbors. Region parameter vectors and the dissimilarity/adjacency matrix are updated at each iteration stage and stop conditions are checked. If no 
regions are merged at the current hierarchy level or the remaining number of regions is below a specified threshold the algorithm stops and the final segmentation is 
returned to the user, otherwise a new iteration is started.

\item \emph{Post-processing}: Some post-processing stages can be performed on the detected sources. A first step uses the hierarchical clustering approach described above to identify 
similar regions within each source and generate a list of nested sources one level down in the source hierarchy. 
Further, following \citet{Yang2008}, a number of statistical and morphology-descriptor parameters are computed over the source contour and/or its pixel distribution to be 
eventually employed in a source classification stage. Standard parameters include bounding box/ellipse, image/contour moments and roundness/rectangularity estimators. 
More complex parameters, such as Fourier Descriptors (FDs) \citep{Zhang2003}, Hu \citep{Hu1962} and Zernike moments \citep{Singh2011}, can be computed and supplied to the user.
\end{enumerate}  

\begin{table*}
\caption{Main parameters used in the source finder algorithm.}
\begin{tabular}{p{1.5cm}|p{1.5cm}|p{8cm}}
\bottomrule%
\textbf{Stage} & %
\textbf{Parameter} & %
\textbf{Description}\\%
\hline%
 \multirow{2}{*}{\pbox{1.5cm}{\texttt{Background}}} & 
 \texttt{bkgModel} & \pbox{8cm}{\vspace{0.1cm}Model to be used for computing the background and noise maps (1=global, 2=local, 3=local robust).\vspace{0.1cm}}\\ \cline{2-3}%
 & \texttt{boxSize} & \pbox{8cm}{\vspace{0.1cm}Size of the box used to compute local background/noise estimators.\vspace{0.1cm}}\\ \cline{2-3}%
 & \texttt{gridSize} & \pbox{8cm}{\vspace{0.1cm}Size of the grid used when interpolating the local background/noise estimators.\vspace{0.1cm}}\\ 
\bottomrule
 \multirow{2}{*}{\pbox{1.5cm}{\texttt{Filtering}}} & \pbox{1.5cm}{\textbf{$\sigma_{\texttt{seed}}$ $\sigma_{\texttt{merge}}$}} & \pbox{8cm}{\vspace{0.1cm}Seed and merge
 threshold used to detect compact 
 bright blobs in the image, e.g. $\sigma_{\texttt{seed}}$=10, $\sigma_{\texttt{merge}}$=2.5.\vspace{0.1cm}}\\ \cline{2-3}%
 & \textbf{$K_{\texttt{dilate}}$} & \pbox{12cm}{Kernel size to be used when dilating bright sources.}\\ \cline{2-3}%
 & \textbf{$\sigma_{\texttt{smooth}}$ $K_{\texttt{smooth}}$} & \pbox{7cm}{\vspace{0.1cm}Kernel and radius parameter to be used in image residual smoothing.\vspace{0.1cm}}\\ \hline%
 \multirow{2}{*}{\pbox{1.5cm}{\texttt{Superpixel}\\ \texttt{Generation}}} & \textbf{$l$} & Superpixel size used to generate the initial superpixel partition.\\ \cline{2-3}%
 & \textbf{$\beta$} & \pbox{8cm}{\vspace{0.1cm}Regularization parameter controlling starting superpixel segmentation and balancing clustering spatial and color distance. 
 Low $\beta$ values favors spatial clustering, high $\beta$ favors color clustering\vspace{0.1cm}}\\ \hline%
 \multirow{2}{*}{\pbox{1.5cm}{\texttt{Saliency}\\ \texttt{Filter}}} & \textbf{$l_{\texttt{min/max/step}}$} & Superpixel sizes to be used in multi-resolution saliency computation, e.g. 
 $l$=20-60, step 10.\\ \cline{2-3}%
 & \textbf{$\texttt{knn}$} & \pbox{8cm}{\vspace{0.1cm}Fraction of nearest neighbors superpixel used in saliency estimation, 
 e.g. $\texttt{knn}$=10\%/20\%\vspace{0.1cm}}\\ \cline{2-3}%
 & \textbf{$f_\texttt{sal}^\texttt{scales}$} & \pbox{8cm}{\vspace{0.1cm}Fraction of salient scales required to contribute to final saliency 
 estimation, e.g. $\texttt{knn}$=70\%\vspace{0.1cm}}\\ \cline{2-3}%
 & \texttt{useCurvMap} & Flag to include (multi-scale) curvature maps in saliency estimation\\ \cline{2-3}%
 & \texttt{useBkgMap} & Flag to include (multi-scale) background map in saliency estimation\\ \cline{2-3}%
 & \texttt{useNoiseMap} & Flag to include (multi-scale) noise map in saliency estimation\\ \cline{2-3}%
 & \texttt{salThrModel} & Method to be used for thresholding final saliency map (1=global, 2=local, 3=local robust)\\ \cline{2-3}%
 & \textbf{$f_{\texttt{thr}}^{\texttt{bkg}}$} & \pbox{8cm}{\vspace{0.1cm}Global threshold parameter to tag background pixel candidates in 
 saliency map, e.g. $f_{\texttt{thr}}^{\texttt{bkg}}$=1.\vspace{0.1cm}}\\ \cline{2-3}%
 & \textbf{$f_{\texttt{thr}}^{\texttt{sig}}$} & \pbox{8cm}{\vspace{0.1cm}Global threshold parameter to tag signal pixel candidates in 
 saliency map, e.g. $f_{\texttt{thr}}^{\texttt{bkg}}$=2.\vspace{0.1cm}}\\ \hline%
 \multirow{2}{*}{\pbox{1.5cm}{
 \texttt{Superpixel}\\\texttt{Merging}}} & \textbf{$\lambda$} & \pbox{8cm}{\vspace{0.1cm}Regularization parameter used in superpixel merging stage balancing appearance and edge 
 terms when computing superpixel dissimilarities. Low $\lambda$ values (close to zero) favors intensity similarity, high $\lambda$ values (close to 1) favors 
 edge penalization.\vspace{0.1cm}}\\ \cline{2-3}%
 & \texttt{Edge Model} & \pbox{8cm}{\vspace{0.1cm}Model to be used to compute superpixel edgeness (1=Kirsch, 2=Chan-Vese).\vspace{0.1cm}}\\ \cline{2-3}%
 & \textbf{$f_{\texttt{merge}}$} & \pbox{8cm}{\vspace{0.1cm}Fraction of top ranked superpixels selected for merging at each hierarchy level, e.g. $f_{\texttt{merge}}$=30\%.\vspace{0.1cm}}\\ \cline{2-3}%
 & $\varepsilon_{\texttt{merge}}^{\texttt{1st,2nd}}$ & \pbox{8cm}{\vspace{0.1cm}Maximum mutual dissimilarity tolerance used for accept a selected superpixel merging for 1st or 2nd 
 neighbor superpixels, e.g. 5-15\%.\vspace{0.1cm}}\\ \cline{2-3}%
 & $\Delta_{\texttt{thr}}$ & \pbox{8cm}{\vspace{0.1cm}Absolute dissimilarity threshold, when applied, to select/reject selected superpixel merging ($\Delta_{ij}\le\Delta_{thr}$). 
 Low $\Delta_{thr}$ values (close to zero) imply strict superpixel similarity for merging. High $\Delta_{thr}$ values relax the merging.\vspace{0.1cm}}\\ \hline%
\end{tabular}
\label{AlgorithmParTable}%
\end{table*}

\subsection{Algorithm implementation}
The described algorithms have been implemented in a C++ software library, dubbed \tool{} (\toolinfo{}), allowing 
image filtering, background estimation, source finding, image segmentation starting from images in FITS or ROOT format. The library is mainly based on the \textit{ROOT} \citep{Brun1997} 
and \textit{R} \citep{R} frameworks for statistical objects and methods and on the \textit{OpenCV} library \citep{Bradski2000} for some of the image filtering algorithms. The source finding and 
segmentation algorithms have been developed from scratch along with some of the employed filtering stages. Future developments include the algorithm fine-tuning and optimization and further design 
activities for ease of deployment in a distributed computing infrastructure and integration within the pipeline frameworks of next-generation telescopes. Public distribution is planned once optimization steps are carried out.

\begin{figure*}
\subfloat[Field B]{\includegraphics[scale=0.23]{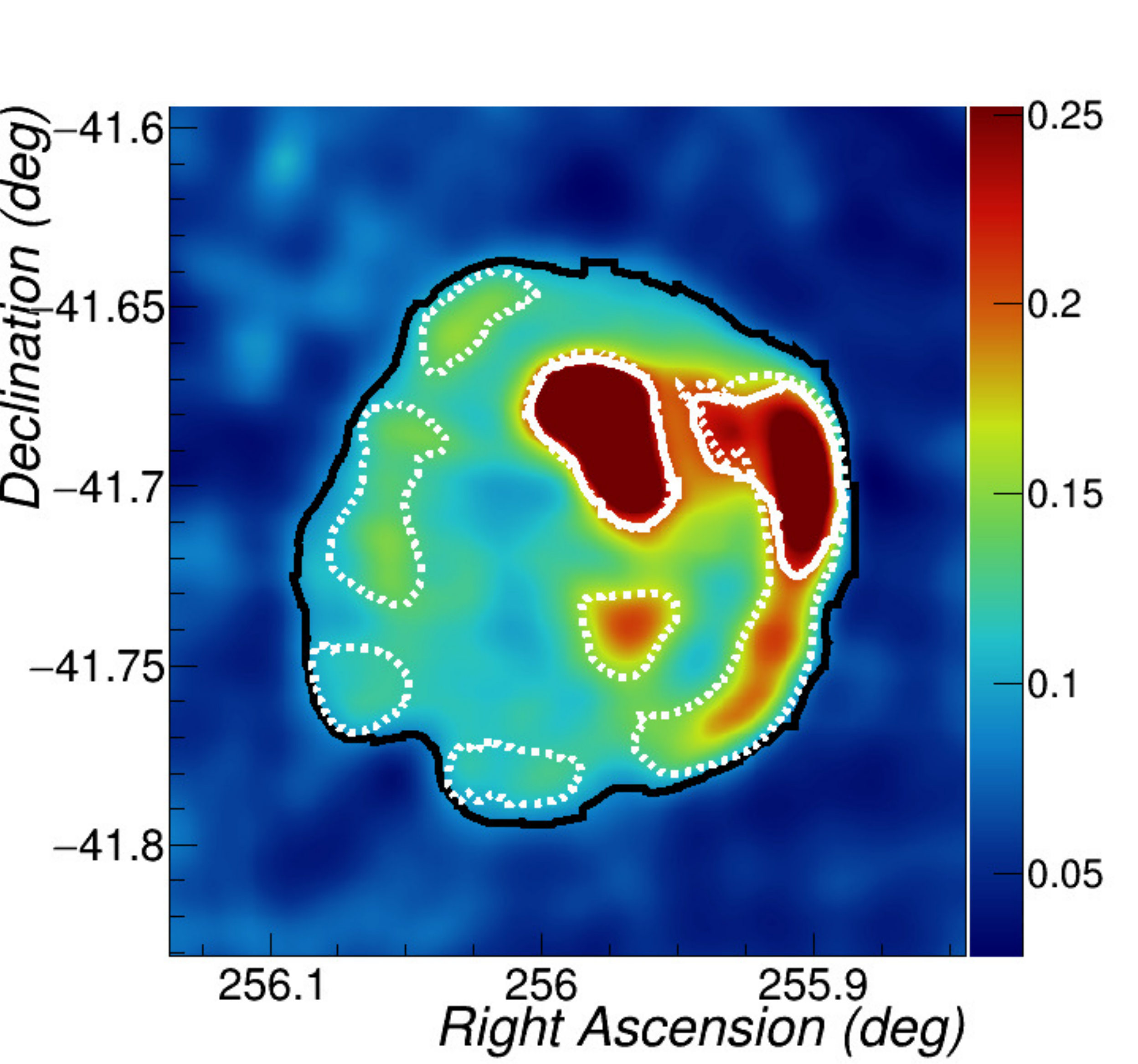}\label{PostProcessingResultsFig1}}
\hspace{-0.2cm}
\subfloat[Field C]{\includegraphics[scale=0.23]{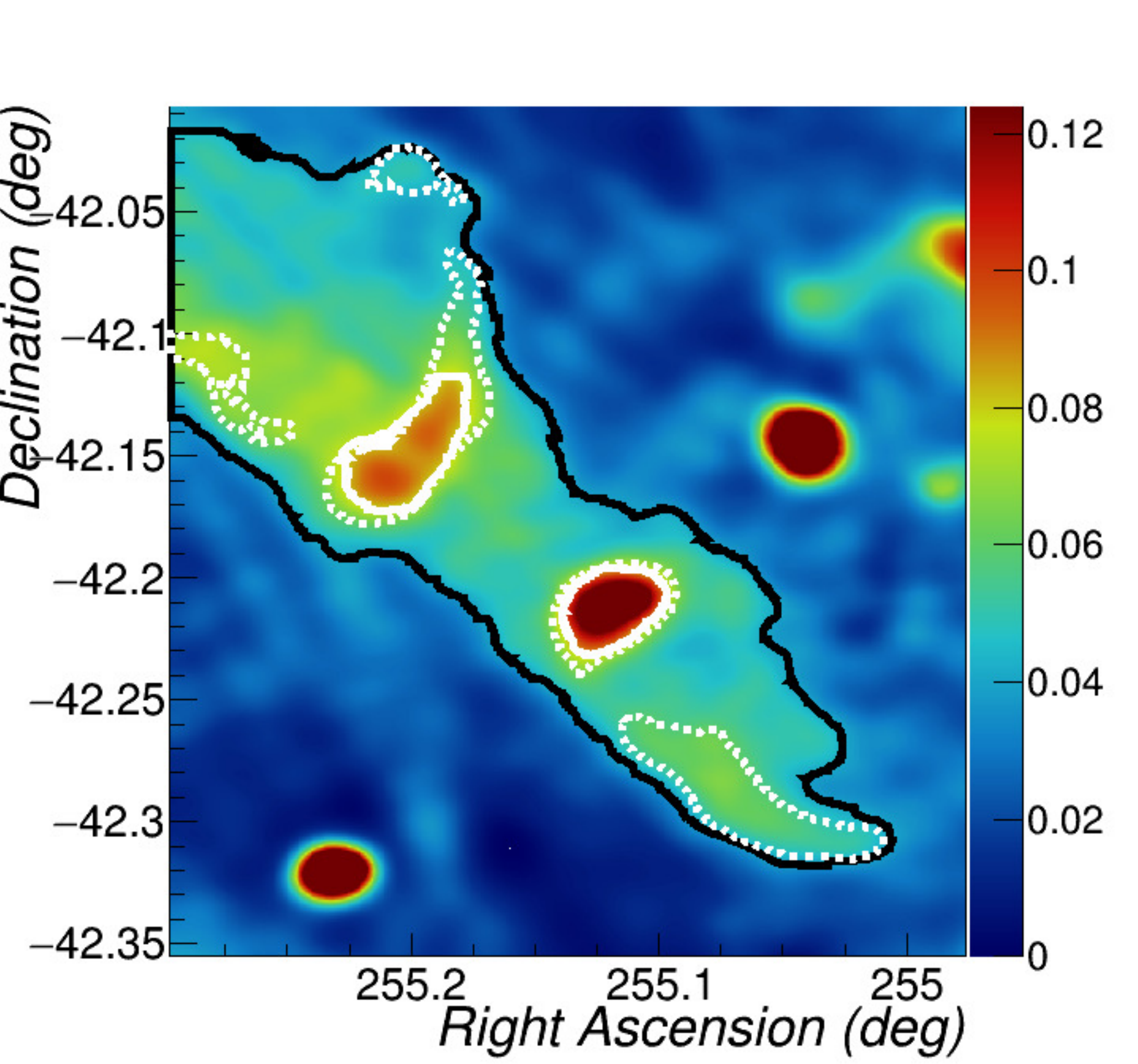}\label{PostProcessingResultsFig2}}
\hspace{-0.2cm}
\subfloat[Field D]{\includegraphics[scale=0.23]{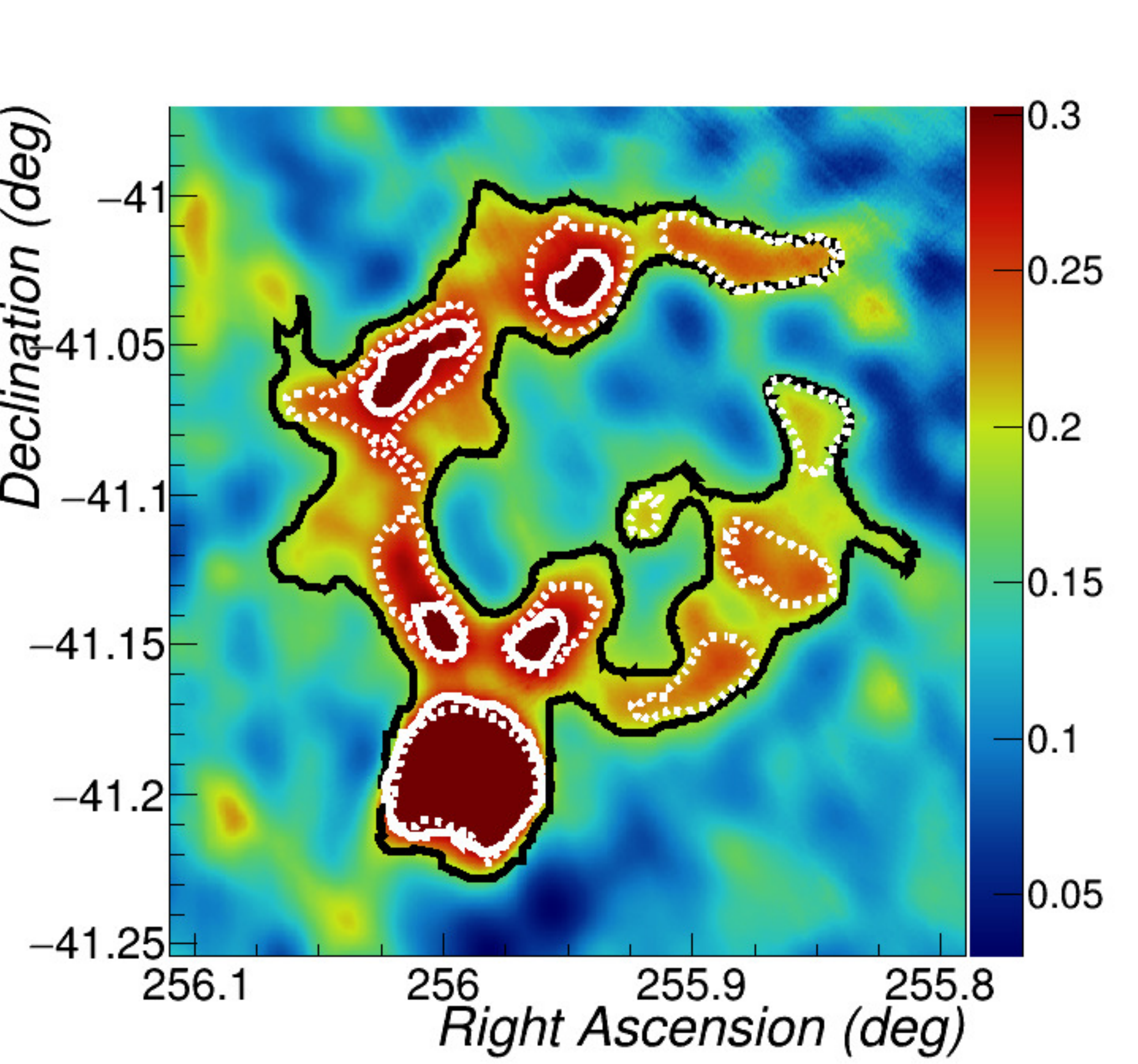}\label{PostProcessingResultsFig3}}\\
\vspace{-0.4cm}
\subfloat{\includegraphics[scale=0.5]{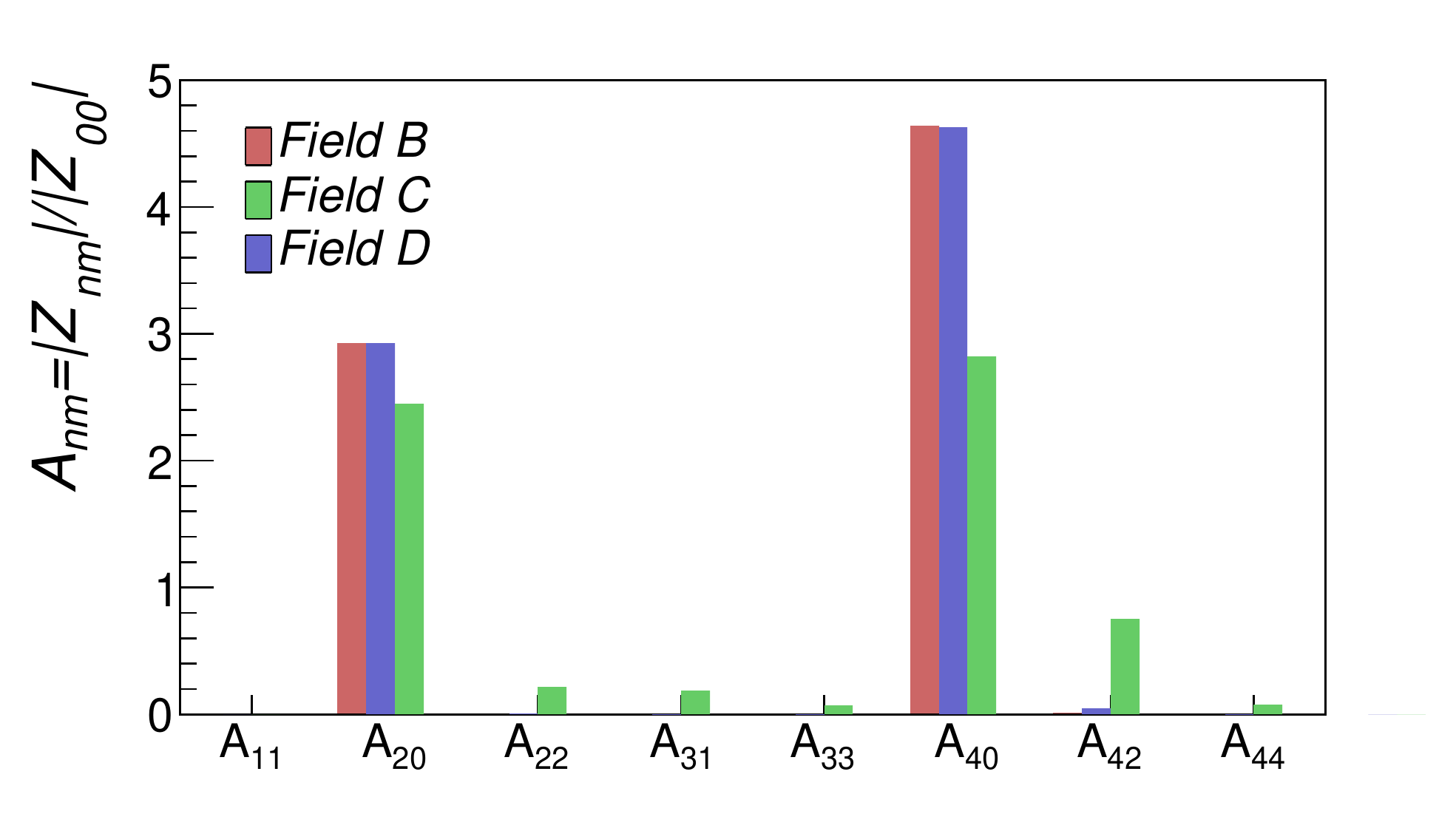}\label{PostProcessingResultsFig4}}
\caption{Top panels: Sample segmented source images, normalized to range [0,1], in field B, C and D (solid black contours). White contours represent nested regions selected 
with a multi-resolution saliency-based method (solid lines) and with a multi-scale blob detector (dashed lines); Bottom panels: Zernike moments up to order $n$=4 computed over the 
segmented sources shown in the upper panels (black contoured area).}
\label{PostProcessingResultsFig}
\end{figure*}

\section{Application to SCORPIO project data}\label{ResultSection}

\subsection{Sample fields}\label{SCORPIOSampleData}
To test the designed algorithm we considered four selected fields from the SCORPIO map in which several extended structures are present 
along with compact sources. The map is built as described in Paper I using data observed with the ATCA 0.75A array configuration in combination with 
data observed with the ATCA EW367 configuration, in which shorter baselines are present. The effective frequency range of the radio data used is 1.4-3.1 GHz. 
The sample fields, hereafter denoted as field A-D are shown in Fig.~\ref{ScorpioFieldFig}, and some details are reported below: 
\begin{itemize}
 \item \emph{Field A} (Fig.~\ref{ScorpioFieldFig1}): Field A (1000$\times$1000 pixels) is centered on the [DBS2003] 176 galactic stellar cluster (l=343.4830$^\circ$, b=-00.0380$^\circ$, 
 angular size=1.45 arcmin). Two bubble objects, S16 and S17 \citep{Churchwell2006}, are associated with the cluster but only S17 is observed in the radio domain. Two bright point-like radio 
 sources (SCORPIO1\_320 and SCORPIO1\_300), already known objects in radio, were identified in Paper I. SCORPIO1\_300 is located within the S17 bubble and has peak 
 flux around 0.04 Jy/beam. The brighter SCORPIO1\_320 (peak flux$\sim$0.14 Jy/beam) has been tentatively classified as a Massive Young Stellar Object (MYSO) candidate \citep{Urquhart2007}.
 \item \emph{Field B} (Fig.~\ref{ScorpioFieldFig2}): Field B (1600$\times$1850 pixels) is centered on the Supernova Remnant (SNR) G344.7-0.1, located in the adjacency of the high 
 energy $\gamma$-ray source HESSJ1702-420 (see \citealt{Giacani2011}). Close to the SNR, in the north-east region of the image, another extended emission is present and most probably 
 associated with the MSC 345.1-0.2 supernova remnant candidate ($l$=345.062, $b$=-0.218 according to the MOST MSC survey at 843 MHz \citep{Whiteoak1996}).
 \item \emph{Field C} (Fig.~\ref{ScorpioFieldFig3}): Field C (1000$\times$1000 pixels) was analyzed in detail in Paper I. Some of the extended regions of emission present were 
 associated with the following IRAS sources: IRAS 16566-4204, IRAS 16573-4214, IRAS 16561-4207. The first is recognized as a massive star formation region, while classification is uncertain for the others.
 \item \emph{Field D} (Fig.~\ref{ScorpioFieldFig4}): Field D (1000$\times$1000 pixels) is centered on the faint SNR Candidate MSC G345.1+0.2. Below this a more intense 
 emission is present, associated with the G345.097+00.136 HII region.
\end{itemize}
An additional control field, free of extended sources and denoted as field E, is considered to study the algorithm response in the absence of any expected signal and tune the 
detection thresholds. Field E is reported in Fig.~\ref{SparseFieldResultsFig} (left panel). This map is built using data observed with the ATCA 0.75A array configuration alone. 
Due to the larger minimum baseline available extended and diffuse sources are strongly filtered out.

\begin{figure*}
\centering
\subfloat[Field A]{\includegraphics[scale=0.22]{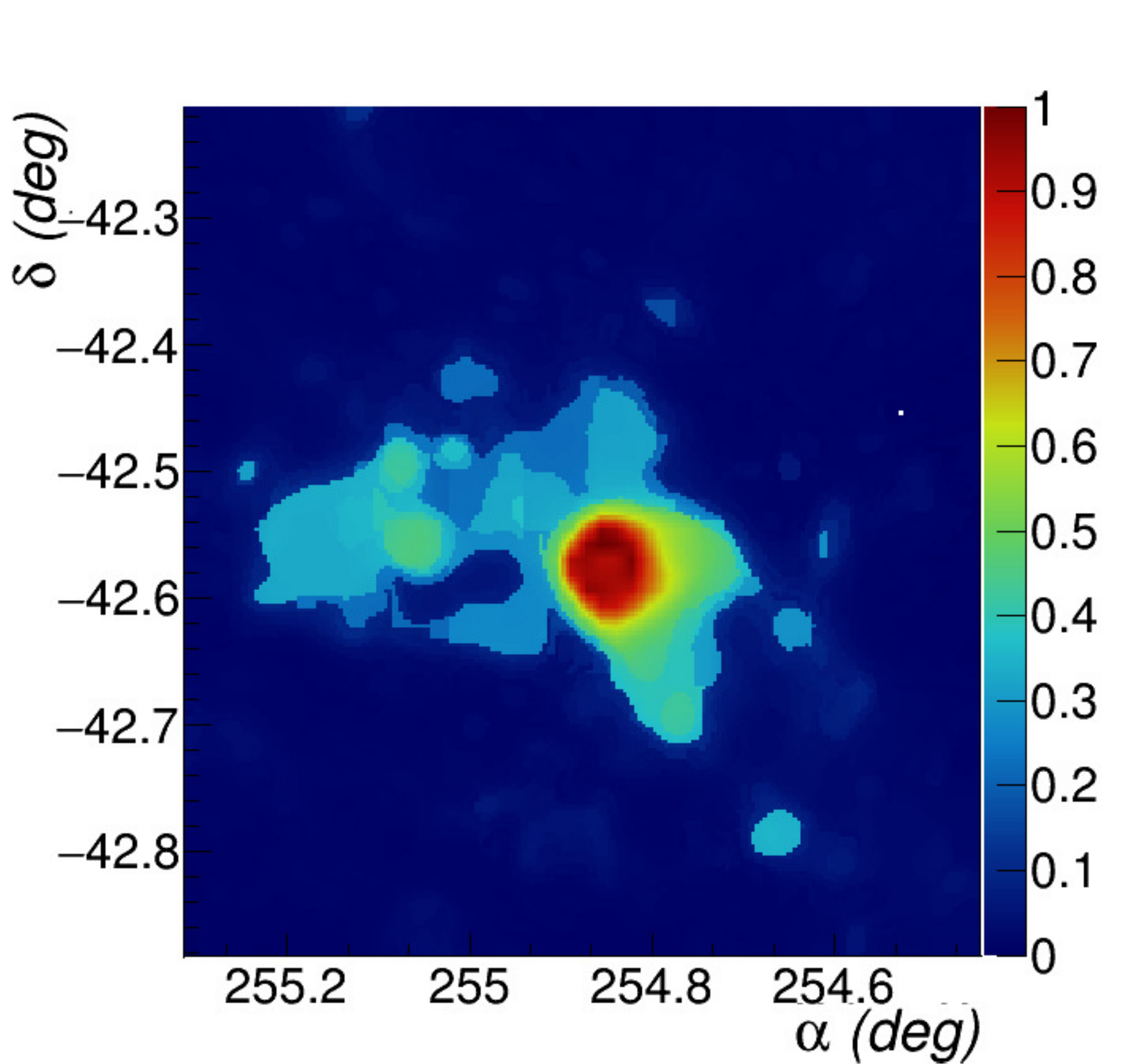}}
\hspace{0.cm}
\subfloat[Field A]{\includegraphics[scale=0.22]{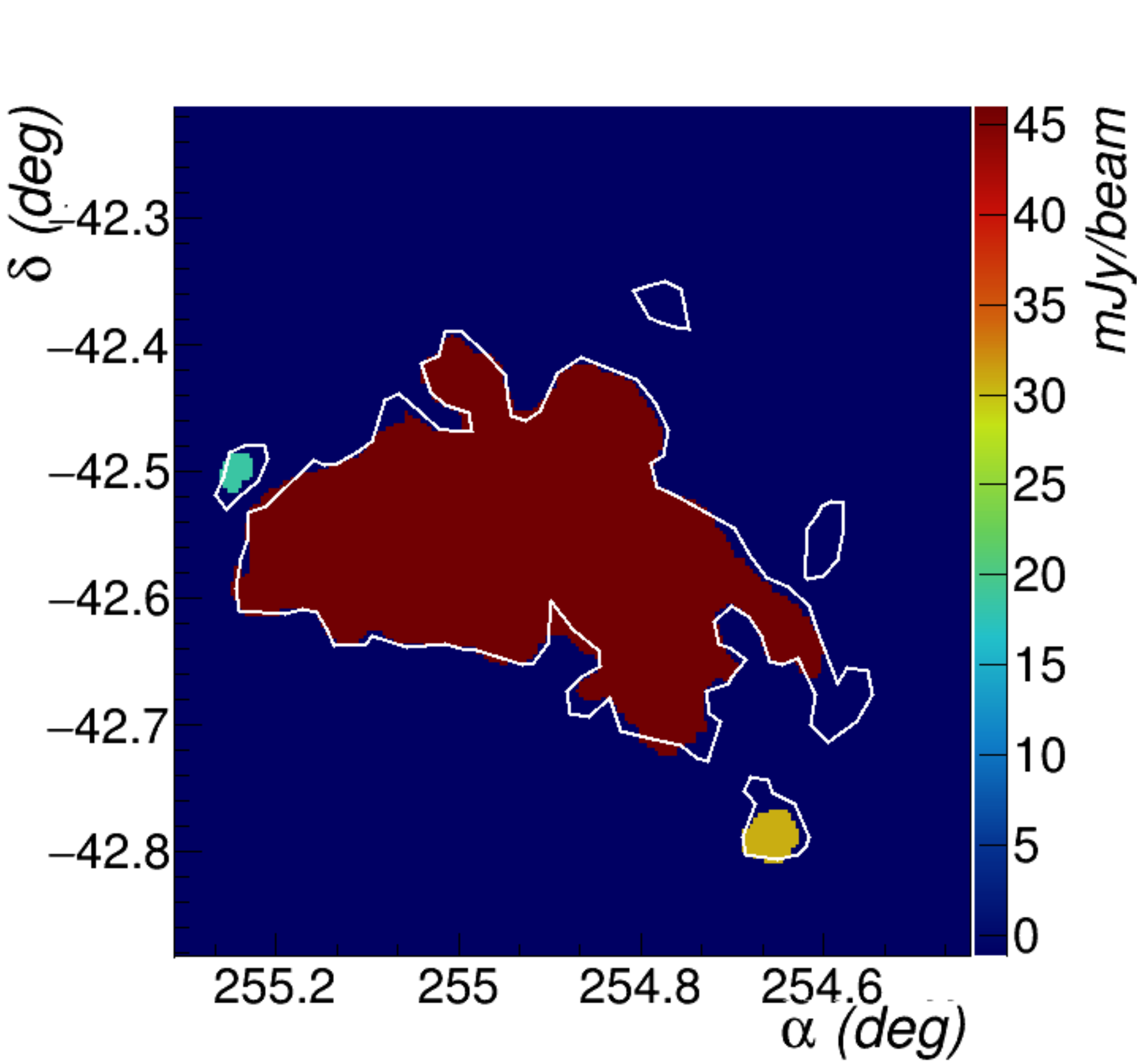}}\\
\vspace{-0.4cm}
\subfloat[Field B]{\includegraphics[scale=0.22]{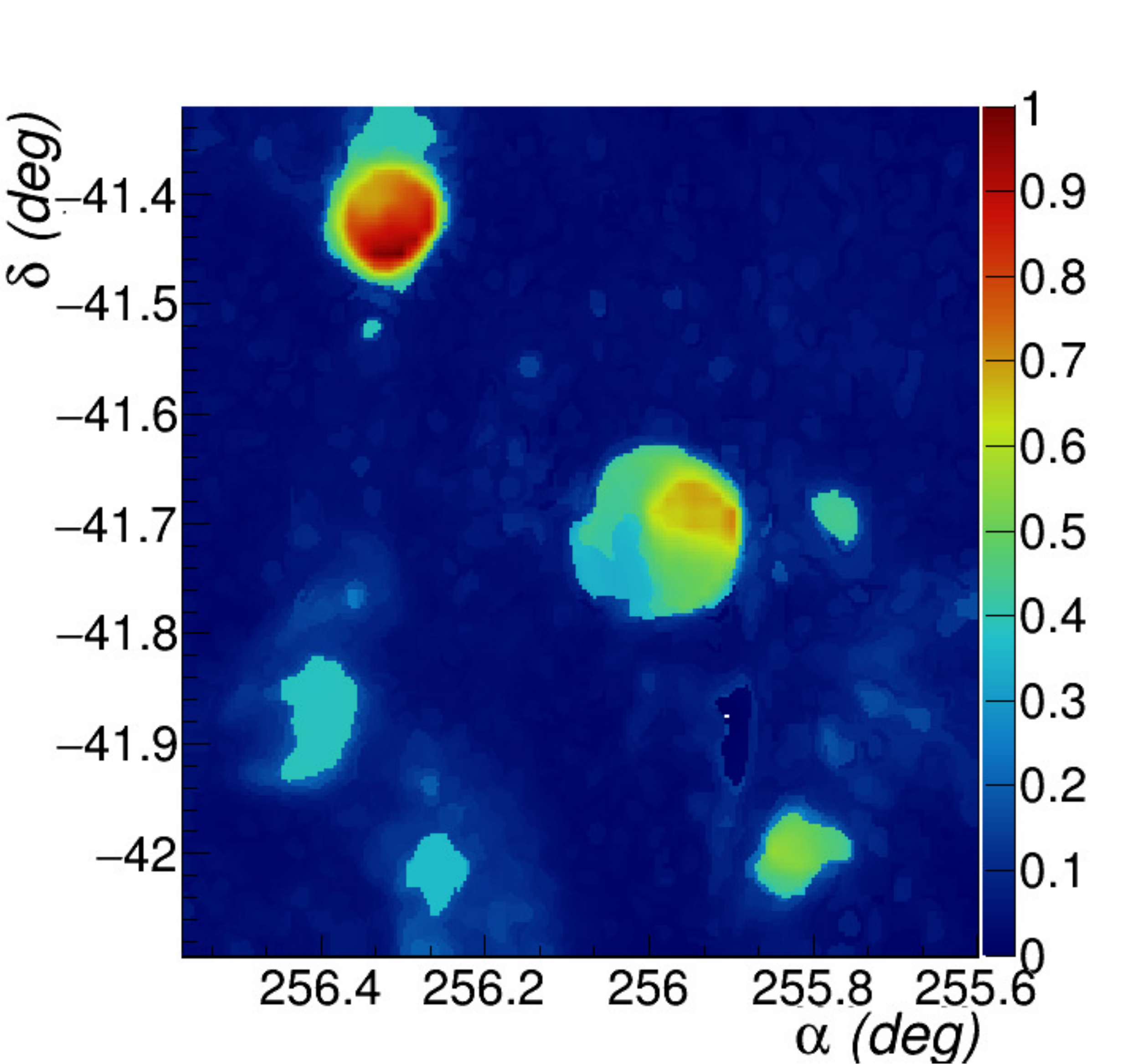}}
\hspace{0.cm}
\subfloat[Field B]{\includegraphics[scale=0.22]{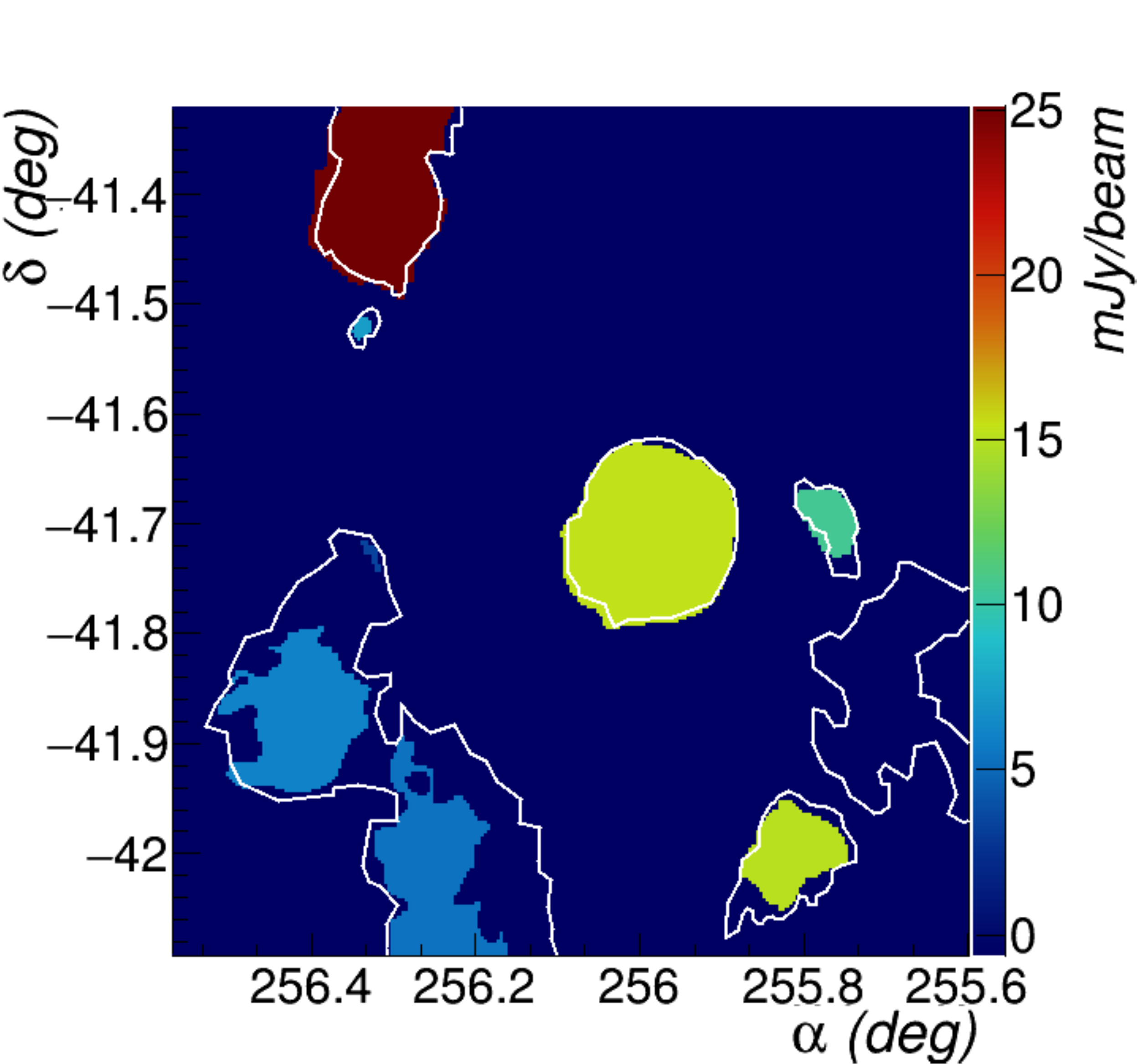}}\\
\vspace{-0.4cm}
\subfloat[Field C]{\includegraphics[scale=0.22]{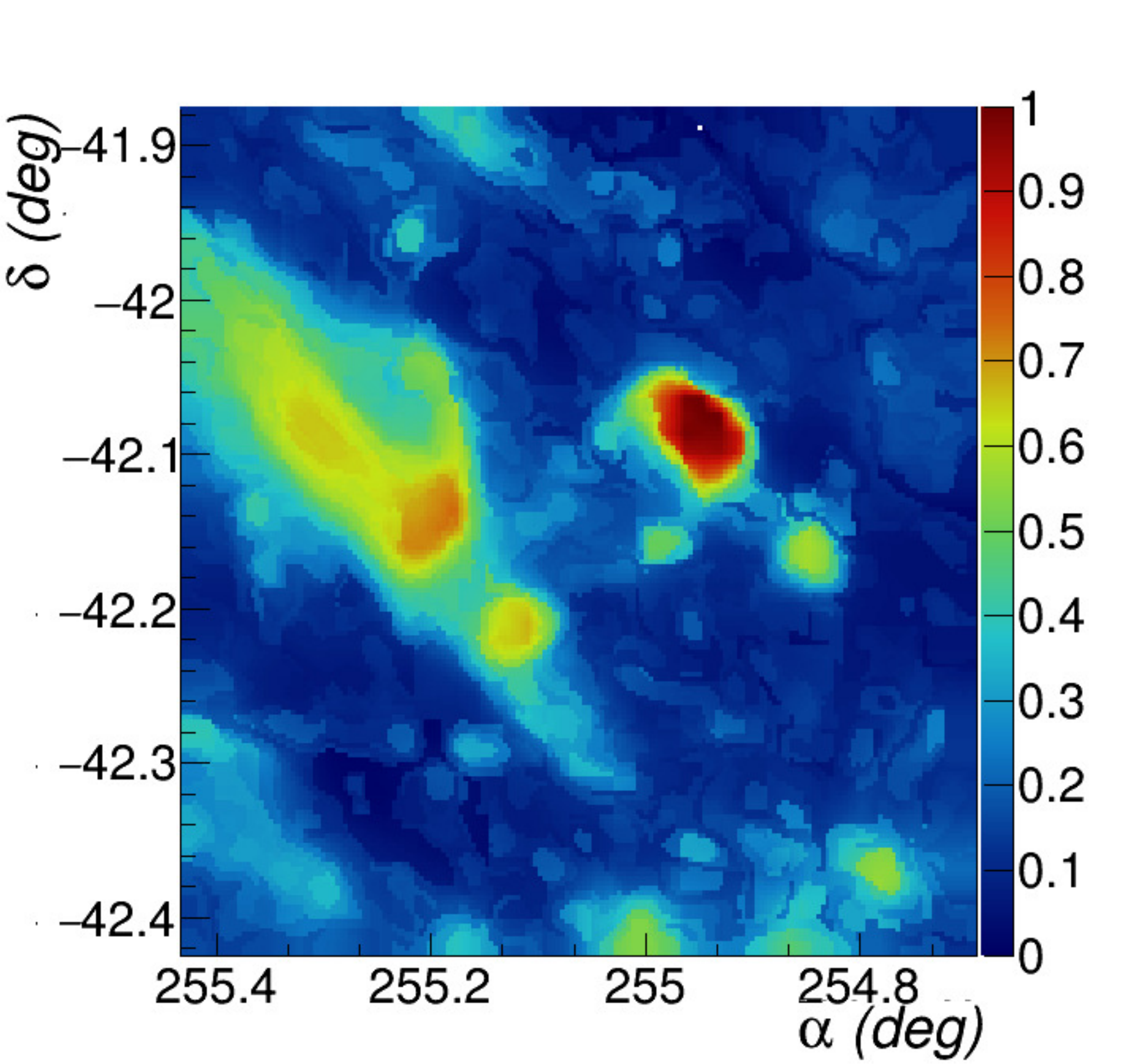}}
\hspace{0.cm}
\subfloat[Field C]{\includegraphics[scale=0.22]{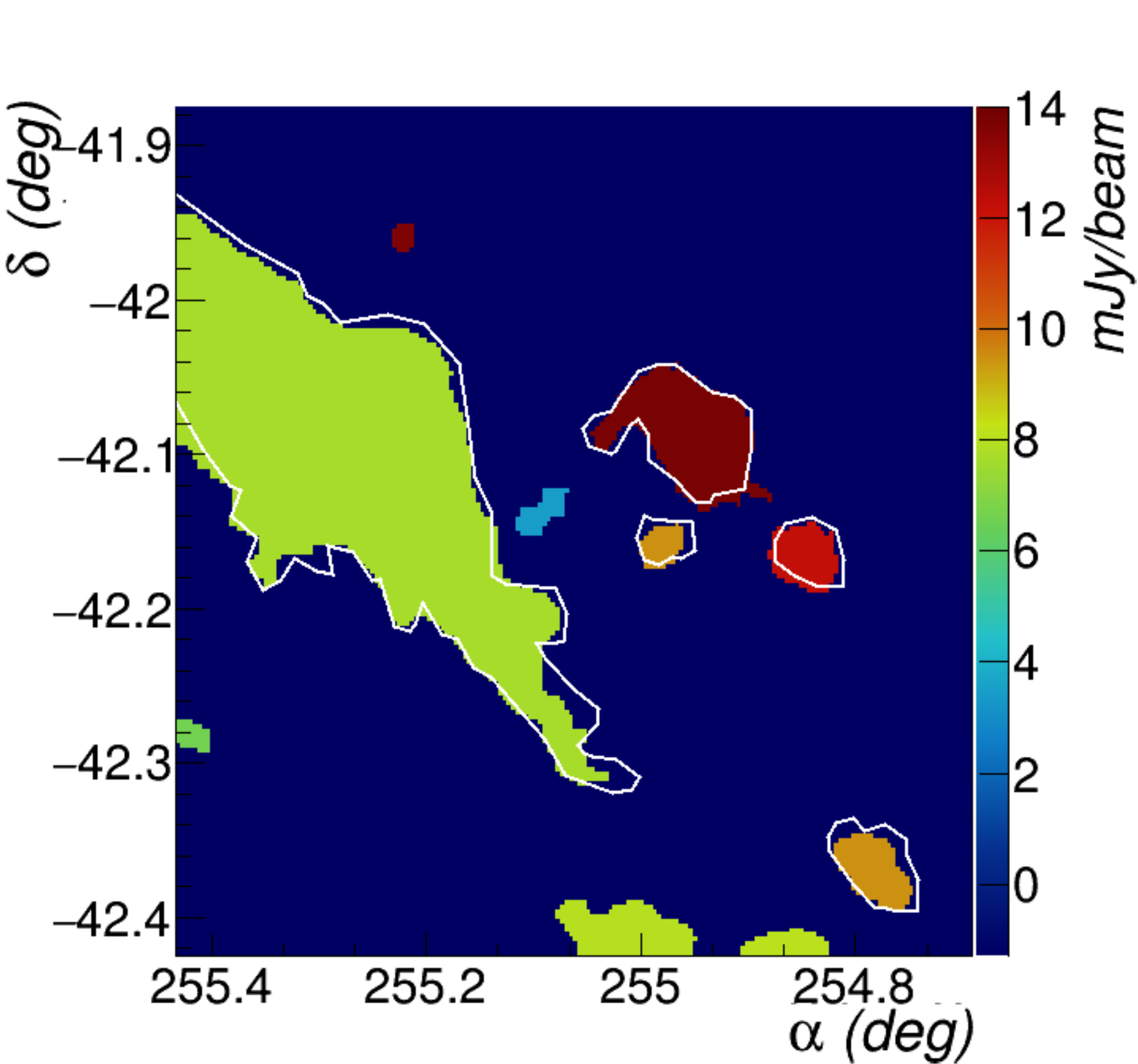}}\\
\vspace{-0.4cm}
\subfloat[Field D]{\includegraphics[scale=0.22]{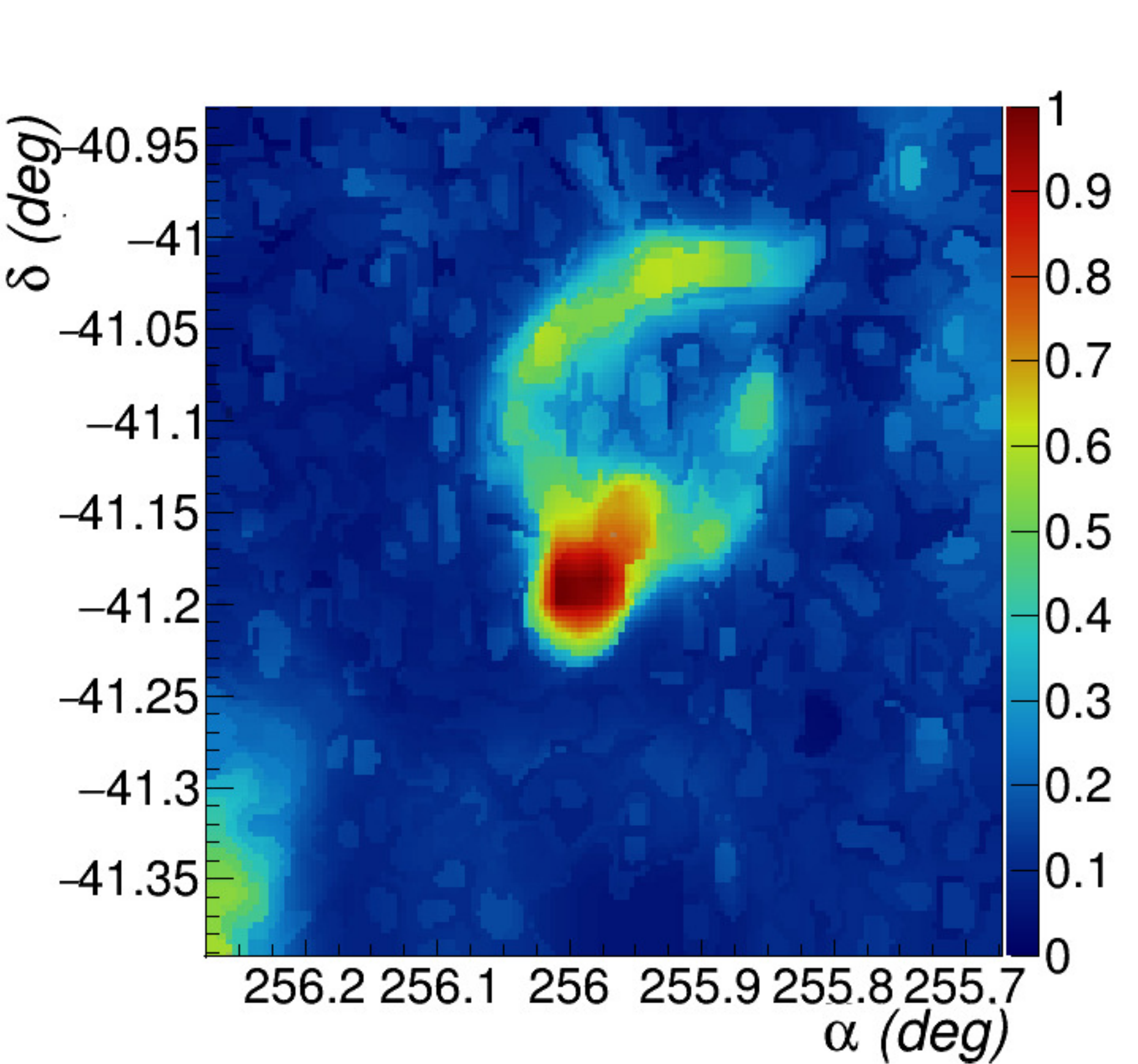}}
\hspace{0.cm}
\subfloat[Field D]{\includegraphics[scale=0.22]{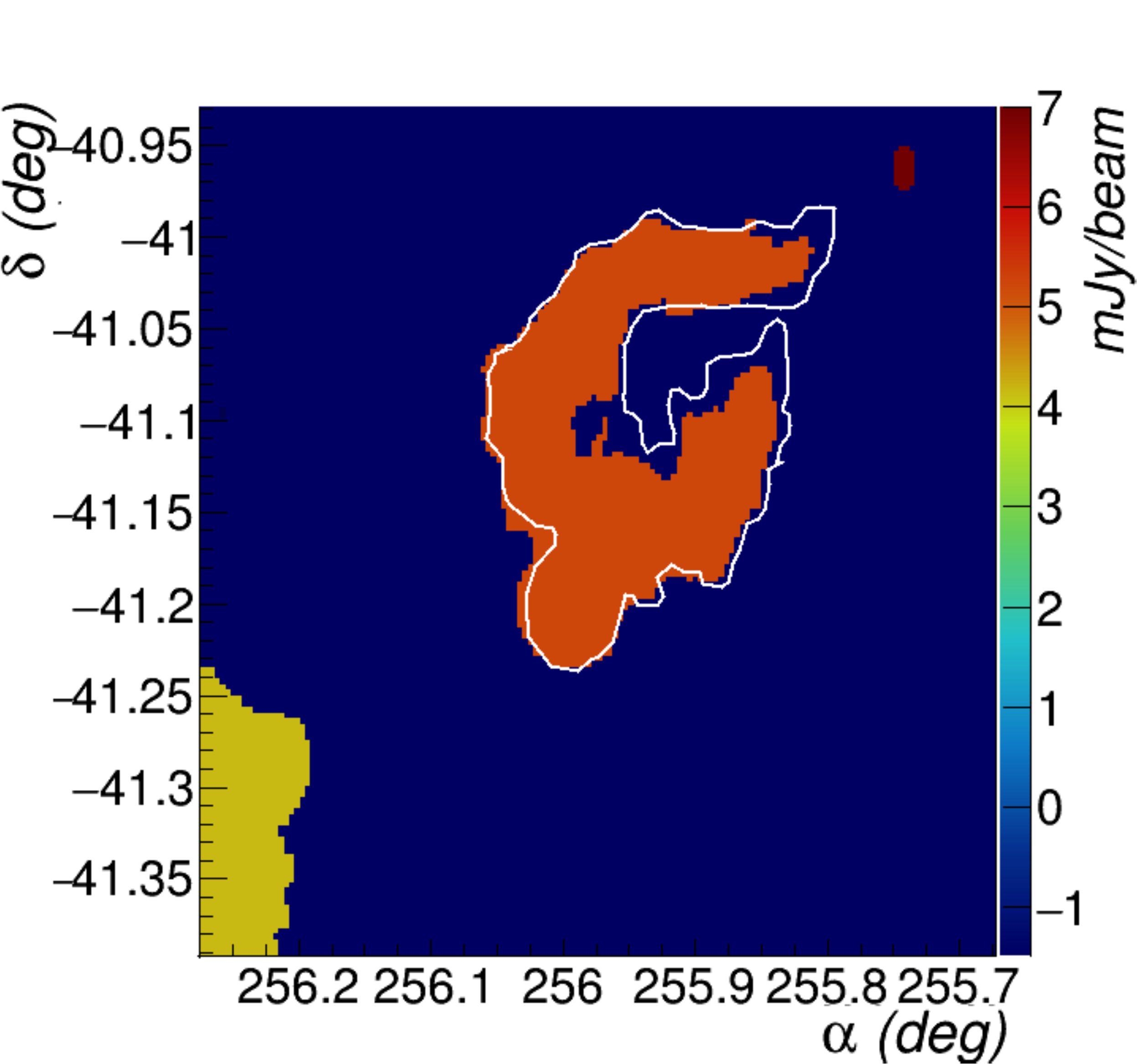}}
\caption{Segmentation results obtained for the Molonglo sample fields A-D (from top to bottom) assuming $l$=5 and $\beta$=1. 
Left: Saliency maps normalized to range [0,1]; Right: Segmentation maps. Each segmented region is colored in the plot according to the mean of its pixel fluxes in mJy/beam units. 
The contours shown with solid white lines correspond to a manual segmentation generated by an expert astronomer.}
\label{MolongloSegmResultsFig}
\end{figure*}
As discussed in Paper I the regions of extended emission present in the test fields A-D are in a few cases firmly associated with real source objects or candidates.
In most cases, however, no association with known sources has been established and an artefact nature cannot be excluded a priori without a further insight and 
comparison to other surveys carried out with different telescopes or wavelength domains. As a result, no ground truth information at pixel level is available to quantify 
the algorithm performances in terms of widely used measures, such as the identification efficiency and false 
detection rate. The quality of the reconstruction will be therefore compared to a human-driven segmentation generated for each sample image by an expert astronomer. To enhance the 
source/artefact discrimination capabilities, we considered the same sample scenarios as observed in the Molonglo Galactic Plane Survey (MGPS) at 843 MHz, reported 
in Fig.~\ref{MolongloScorpioFieldFig}. The rms sensitivity over the survey is around 1-2 mJy/beam and the positional accuracy is 1-2". The lower resolution appears evident, particularly 
in Field B and C in which some of the extended regions present in SCORPIO are not fully resolved and are detected as compact sources in the source finding stage. On the other hand, due 
to the lower observing frequency, regions of extended emission are brighter and can be detected at higher significance levels. Furthermore, it is 
unlikely that the same imaging artefacts appear in both surveys which are conducted with different telescopes. Thus, common emission features can be therefore 
considered as real with a high degree of confidence.

\begin{figure*}
\subfloat[Field B - \textit{Aegean}, \textit{Blobcat}]{\includegraphics[scale=0.27]{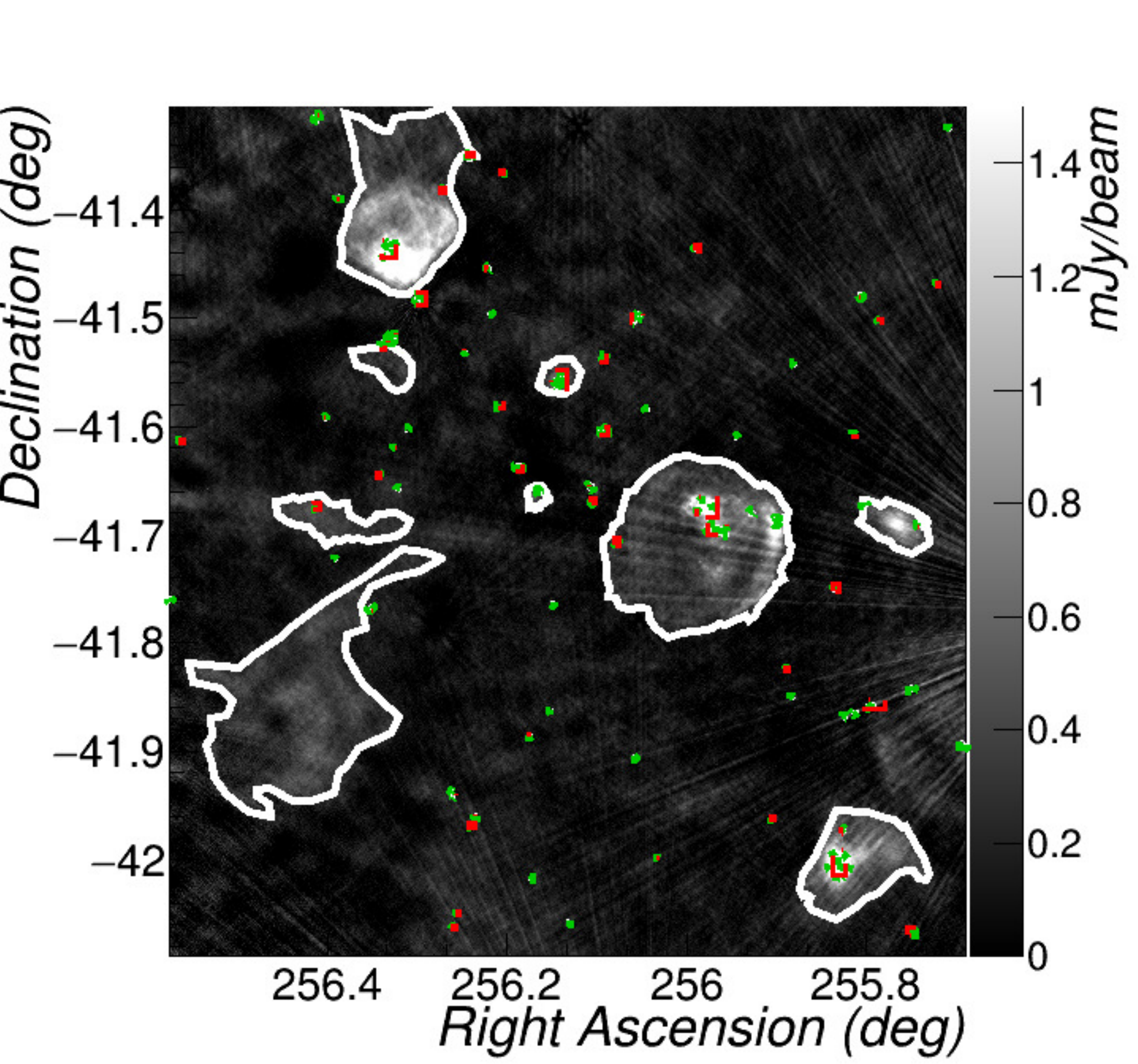}\label{OtherMethodResultsFig1}}
\hspace{-0.1cm}
\subfloat[Field D - \textit{Aegean}, \textit{Blobcat}]{\includegraphics[scale=0.27]{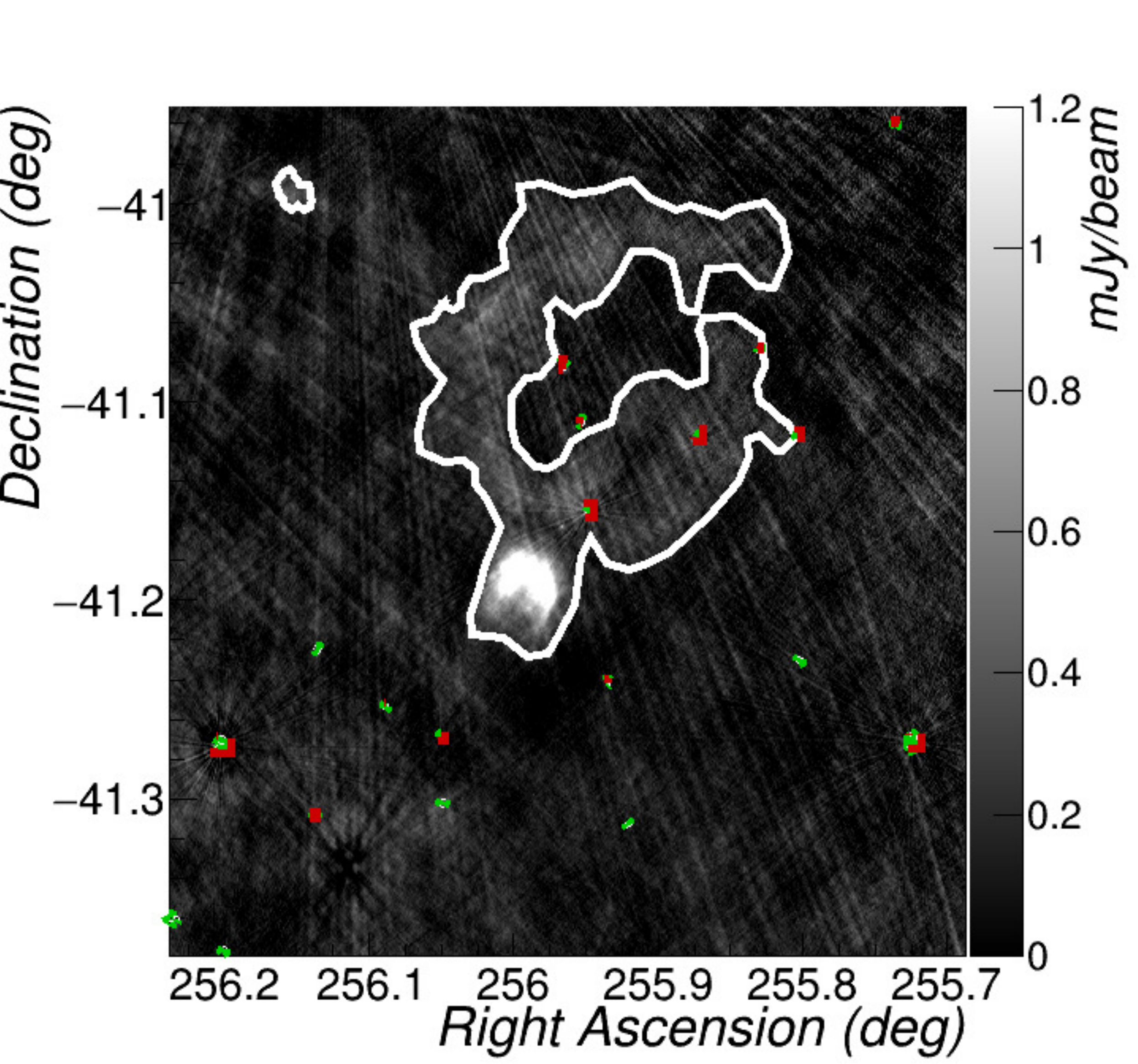}\label{OtherMethodResultsFig4}}\\
\vspace{-0.4cm}
\subfloat[Field B - \textit{Chan-Vese}]{\includegraphics[scale=0.27]{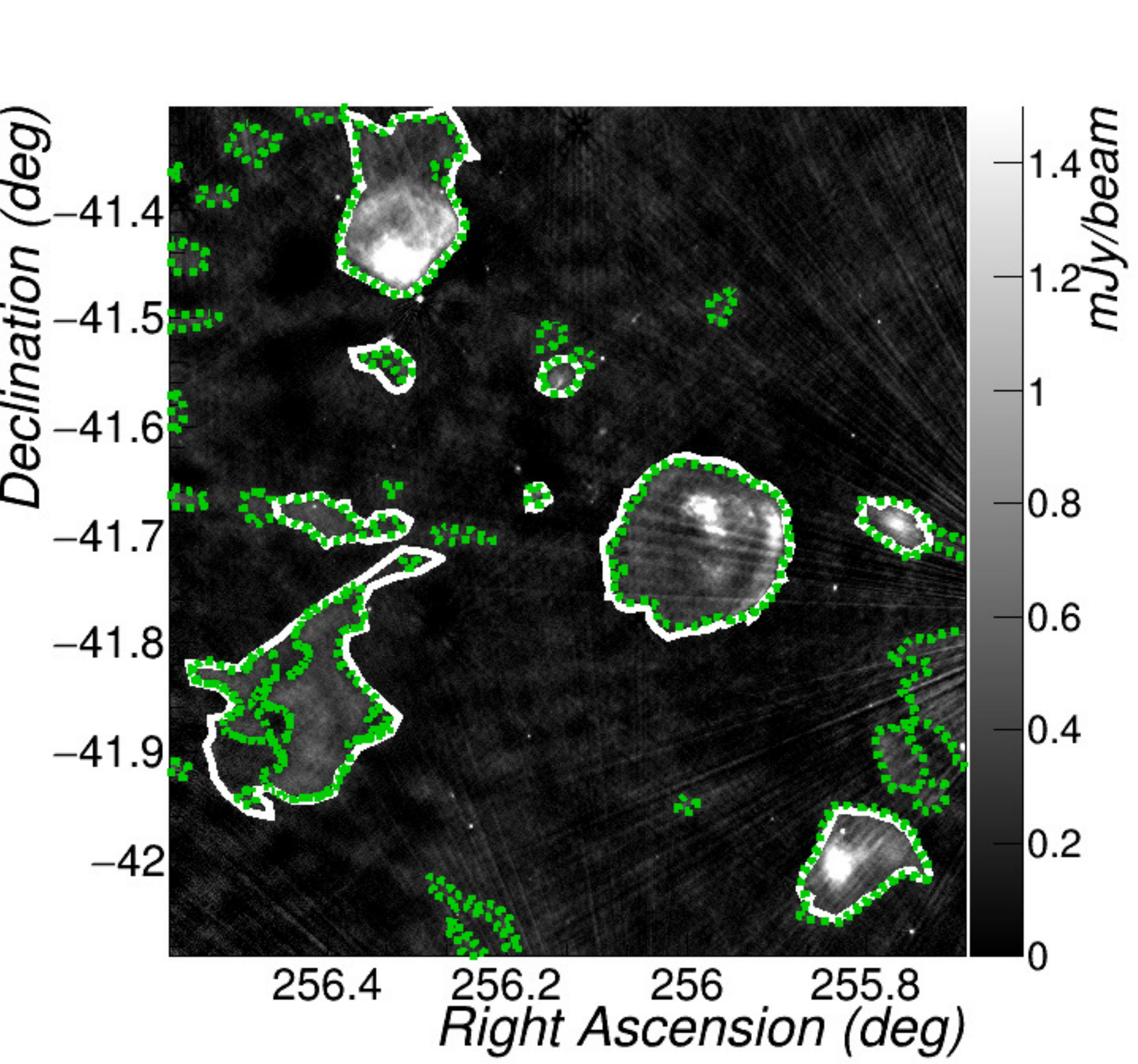}\label{OtherMethodResultsFig2}}
\hspace{-0.1cm}
\subfloat[Field D - \textit{Chan-Vese}]{\includegraphics[scale=0.27]{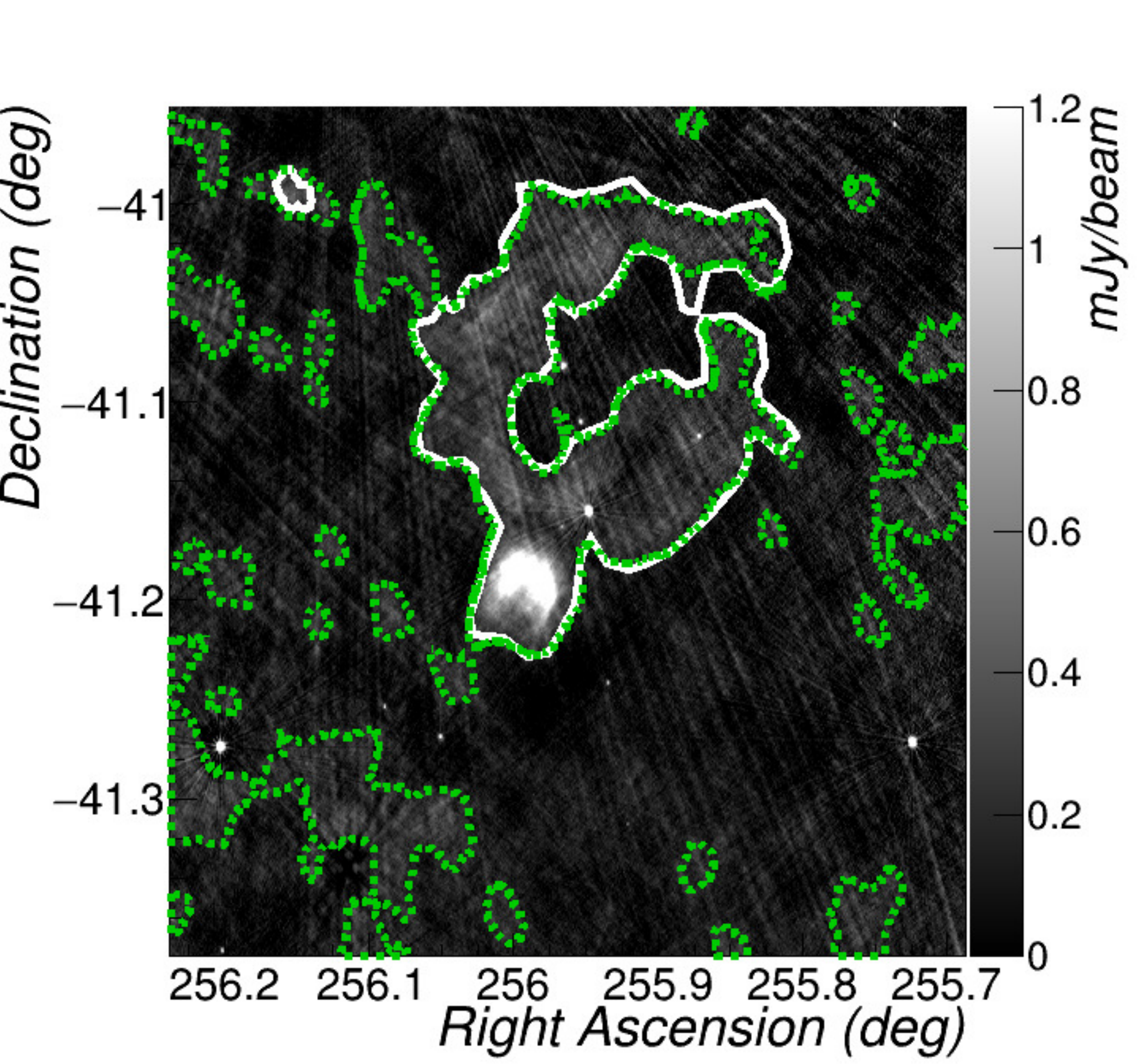}\label{OtherMethodResultsFig5}}\\
\vspace{-0.4cm}
\subfloat[Field B - \textit{SWT}]{\includegraphics[scale=0.27]{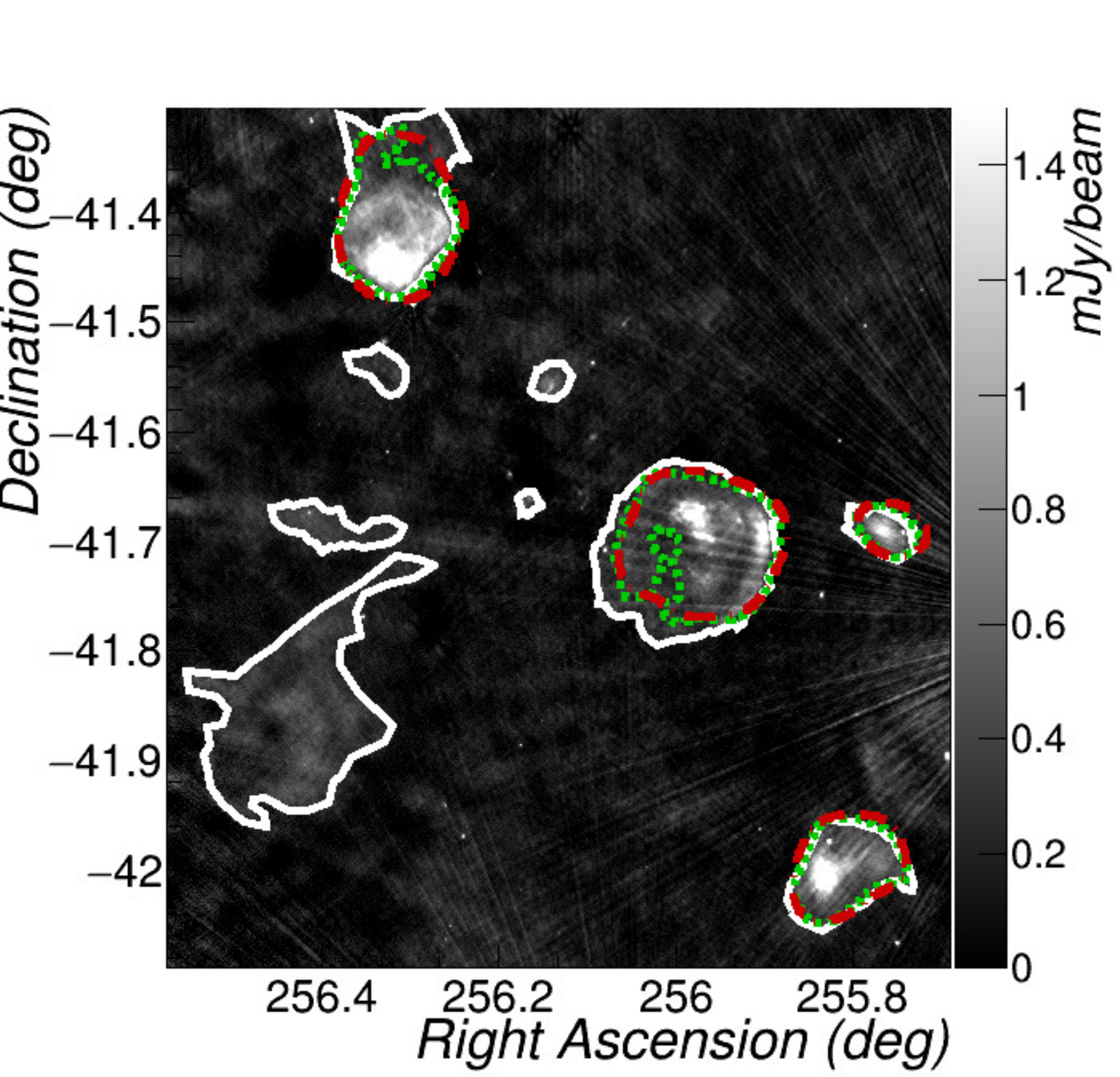}\label{OtherMethodResultsFig3}}
\hspace{-0.1cm}
\subfloat[Field D - \textit{SWT}]{\includegraphics[scale=0.27]{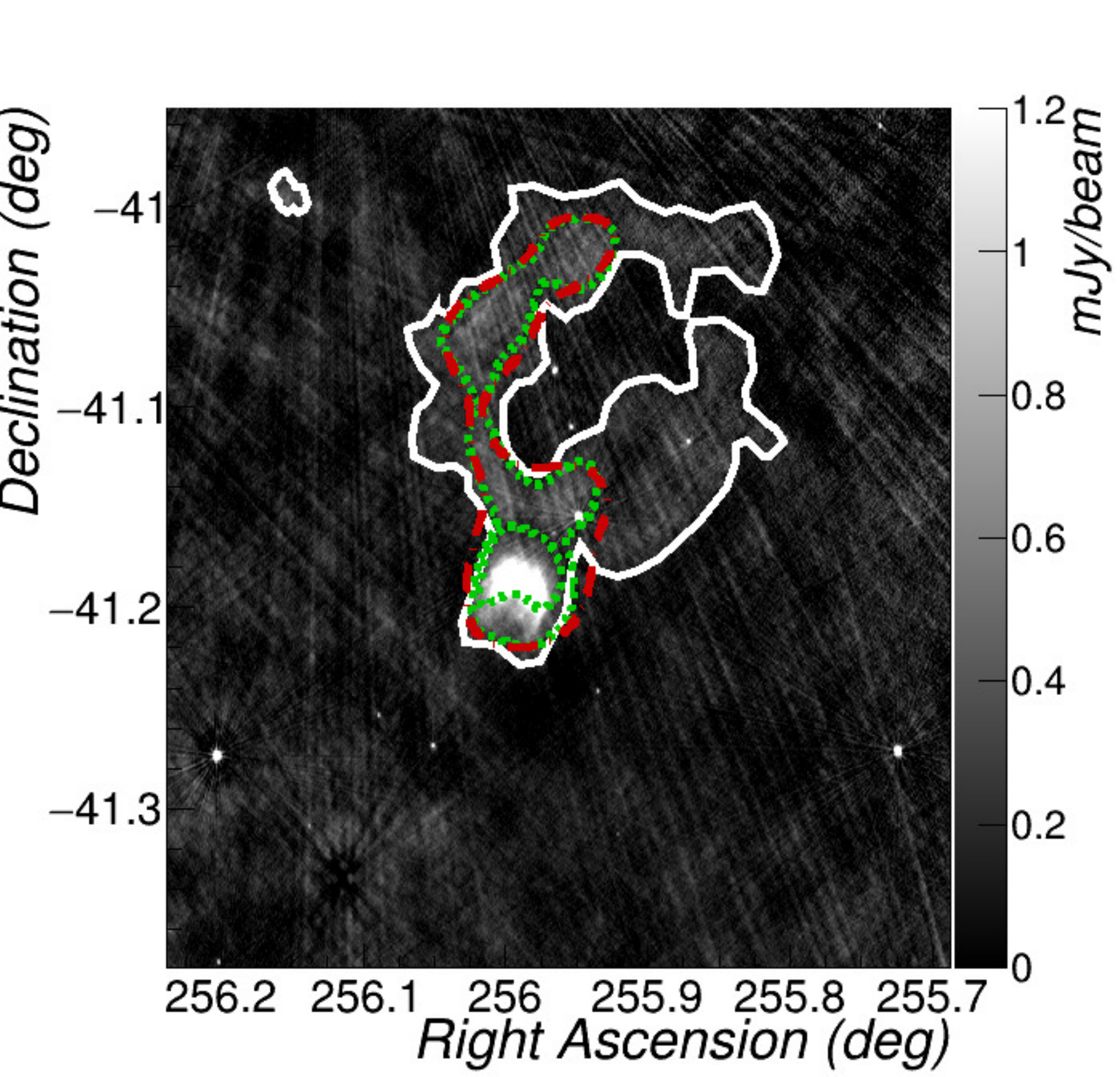}\label{OtherMethodResultsFig6}}
\caption{Source finding results obtained with three different algorithms over field B (left panels) and field D (right panels) compared to the human segmentation 
(solid white contours); Top: Results obtained with the Aegean (dotted green contours) and Blobcat source finders (dashed red contours); Center: Results obtained with the Chan-Vese algorithm 
(dotted green contours); Right: Results obtained with the Stationary Wavelet Transform (SWT) method at scale $J$=5 (dotted green contours) and $J$=6 (dashed red contours).}
\label{OtherMethodResultsFig}
\end{figure*}

\subsection{Results}
We applied the designed segmentation algorithm to the selected test fields described in Section~\ref{SCORPIOSampleData}. 
Multiple runs were performed under different choices of the algorithm parameters. The quality of the segmentation was visually inspected against
the human segmentation and a suitable choice of the algorithm parameters selected on the basis of the maximum 
number of expected objects detected in all test fields at the corresponding minimum false detection rate.

A minimum region size $l$ for the initial segmentation equal to $l\sim4\times$\texttt{beam} (equivalent to $l$=20 pixels) was considered. Smaller values (e.g. $l$=10 pixels), comparable to the 
beam size, were found to be too sensitive to small-scale structures (residual compact emission, artefacts) in the image and thus provide noisy segmentation results.
Larger values, e.g. $l$=30-60 pixels, were investigated as well. As $l$ increases, small-scale details of the extended sources may be smoothed out. This does not represent an issue 
for field A and B in which the extended emission scale is larger by a factor of 4-5 compared to the minimum region size. Furthermore a larger value of $l$ favors the 
merging of artefacts in the background region, e.g. in Field B. 

The regularization parameter $\beta$, controlling initial over-segmentation, was studied. Different values were considered ($\beta$=0.01, 1, 10, 100) in correspondence to 
all other scanned parameters. Results were found comparable for $\beta$=0.01-1 while for values above $\beta$=10 the superpixels start to assume very compact shapes and does 
not fit well to object boundaries.

The saliency maps computed for the SCORPIO sample fields using a multi-resolution range of $l$=20-60 pixels (step 10 pixels), in combination with background and noise maps, are shown 
in the left panels of Fig.~\ref{ScorpioSegmResultsFig}. It can be noted how the faint diffuse emission, previously hardly detectable without manually adjusting the map contrast, is 
significantly enhanced over the background after the saliency filter. The filter mostly preserves the expected object contours and slightly smooth out small scale details.
A thresholding procedure on these saliency maps provides the initial signal and background markers for the following algorithm stages. Suitable values of the global signal 
threshold factor $f_{\texttt{thr}}^{\texttt{sig}}$ were searched over all test samples. The choice of the threshold level was mainly driven by Field D and control Field E and 
optimal values were found in the range 2.5-2.8. Higher values (up to 3.0) can be given to other fields at the cost of missing parts of the faint SNR source in Field D and of the 
large diffuse emission in Field C. Overall, we have found that the thresholded saliency map alone already provides a reasonable source detection. It is also worth to observe 
that saliency maps may constitute a valid input for different algorithms.

Different choices of the similarity regularization parameter $\lambda$ were investigated: $\lambda$= 0, 0.1, 0.5. Results obtained with $\lambda$=0.1, 0.5 are overall comparable, with 
slightly better results obtained with $\lambda$=0.5, while worse results are obtained with $\lambda$=0. This analysis demonstrates that incorporating an edge information in the algorithm
improves the segmentation quality, even though edges of radio objects are considerably softer than in natural images.

The results of the segmentation stage are reported in the right panels of Fig.~\ref{ScorpioSegmResultsFig} for the four tested fields assuming $l$=20 pixels, $\beta$=1 and 
$\lambda$=0.5. Each segmented region is colored according to the mean of its pixel fluxes. The human segmentation is superimposed and shown with solid white
contours. As it can be seen known objects and regions of diffuse emission are all identified and kept for later post-processing. The algorithm, at least with this choice 
of parameters, is also sensitive to other faint diffuse emission which were not identified in the human segmentation. 
After a deeper inspection, some of these were clearly attributed to imaging artefacts present in the input map, particularly 
in the field B in which a poorly cleaned bright object outside the studied field pollutes the entire map. For the remaining objects the nature remains unclear even after a visual 
inspection. This kind of artefacts represents a limitation in current SCORPIO map release. They can be removed in our analysis by increasing the threshold levels in the saliency map, at the
cost of affecting source detection especially in fields C and D. 

In Fig.~\ref{SparseFieldResultsFig} (right panel) we report the results obtained over test field E using the same algorithm parameters selected for fields A-D. The left panel 
shows the input map while the right panel the map given to the segmentation algorithm after the compact source filtering and smoothing stage. As desired, no signal markers are 
found in the saliency map and thus no extended source detection is reported.

An example of post-processing analysis, carried out for some relevant sources present in the test fields, is reported in Fig.~\ref{PostProcessingResultsFig}. Top panels 
shows the identified sources (solid black line contours) with nested components detected using two different methods. Solid white line contours are obtained by 
thresholding a multi-resolution saliency map computed over source pixels. Dashed white line contours are produced by a multi-scale blob detector approach, combining Laplacian 
of Gaussian (LoG) image filters at different scales. Other analysis are possible with the designed algorithm, e.g. running the hierarchical clustering over the source 
region to identify the most similar areas, thus not shown here. 

As discussed in Section~\ref{SegmentationAlgorithmSection} a set of parameters can be computed for each detected source, even the nested ones. As an example we report in the 
bottom panel of Fig.~\ref{PostProcessingResultsFig} the set of Zernike moments computed for the three sources up to the 4-th order. Note how the moments are sensitive to the 
source morphology and can be in principle considered for classification studies in combination with the other computed parameters (not in this paper purposes). A study of the 
suitable set of parameters and their robustness to noise is planned to be performed using simulated data.

\begin{figure}
\centering%
\includegraphics[scale=0.38]{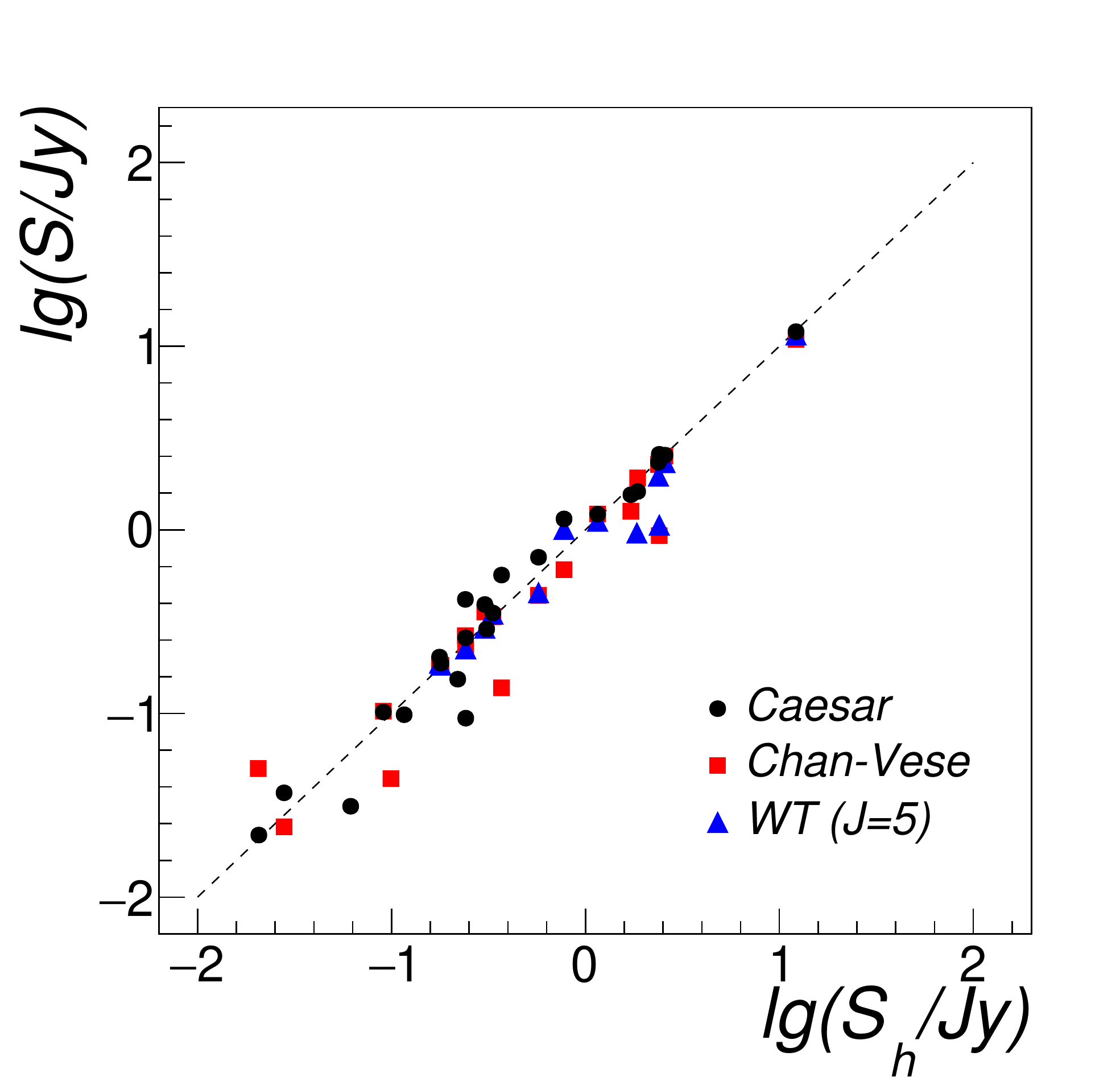}
\caption{Integrated fluxes $S$ of extended sources in the test fields A-D, reconstructed with three different algorithms (black dots: \tool, red squares: Chan-Vese, blue triangles: Wavelet 
Transform J=5), as a function of the human-driven segmentation flux $S_{h}$.}
\label{SourceFluxComparisonFig}
\end{figure}

\subsection{Application to data at different wavelengths}
To evaluate the results obtained on radio data collected at different wavelengths and detector resolutions/sensitivities we considered the same test scenarios as observed in the 
Molonglo Galactic Plane Survey (MGPS) at 843 MHz, shown in Fig.~\ref{MolongloScorpioFieldFig}. We applied our method to the sample Molonglo fields using the same
parameters considered in the analysis of the SCORPIO fields, with the following exceptions related to the lower resolution and size of the Molonglo maps. Smaller values of the 
superpixel sizes ($l$=5-10 pixels) can be 
assumed with respect to the SCORPIO maps, in which we have considered a minimum value of $l$=20 pixels. Saliency maps have been therefore computed starting from the chosen minimum superpixel
size up to a smaller maximum scale value compared to that assumed in SCORPIO maps. A less aggressive initial smoothing filter is also assumed in this case. All the other algorithm 
parameters are left unchanged. The results are reported in Fig.~\ref{MolongloSegmResultsFig}. Some of the extended sources present in the field are not resolved and are
detected as compact sources in the pre-filtering stage. The white contours shown in the plots are therefore relative to the detectable extended sources.
As it can be seen, all the known sources are detected with high fidelity when comparing to the superimposed human segmentation. Additional regions of diffuse emission are detected as well.
It is unclear at the present status whether they are real or most probably reconstruction artefacts. Overall, the results demonstrate that the method is flexible to be used 
also with different data under a minor tuning of parameters driven by the data itself, mainly sensitivity and resolution. 

\subsection{Results with different algorithms}\label{ResultsDifferentAlgorithms}
It is valuable to consider what can be achieved on SCORPIO observed fields with other existing algorithms. 
Such a test is indeed useful to be carried out as many of the available algorithms were tested with less-sensitive radio data or benchmarked against simulated data neglecting the 
real background behavior and the Galactic Plane diffuse emission. 

Four different methods were considered and tested. The first two, \textit{Aegean} \citep{Hancock2012} and \textit{Blobcat} \citep{Hales2012} use a flood-fill
algorithm to detect blobs in the image, starting from pixels above a seed threshold $\sigma_{\texttt{seed}}$ ($\sigma_{\texttt{seed}}$=5) with respect to the background and 
aggregating adjacent pixels above a second lower threshold $\sigma_{\texttt{merge}}$ ($\sigma_{\texttt{merge}}$=2.6). Blobs are finally deblended using curvature information. 
Background and noise maps were computed using the \textit{BANE} tool distributed within the \textit{Aegean} source finder. A third method, adopted by \citet{Peracaula2011}, searches 
for blobs on the Stationary Wavelet Transform (SWT) of a residual image, obtained from the input map by replacing bright compact sources with a random background estimate. We implemented 
this method from scratch. Finally an implementation of the Chan-Vese active contour algorithm \citep{ChanVese2001} was considered and tested over the sample data. The method 
iteratively evolves an initial contour till convergence on the boundaries of the foreground region. Contour evolution is done by seeking a level set function that minimizes a fitting energy 
functional depending on a set of input parameters.

In Fig.~\ref{OtherMethodResultsFig} we report the sources detected by the four methods (from top to bottom) in fields B (left panels) and D (right panels) in comparison with 
the human segmentation shown with solid white contours. \textit{Aegean} and \textit{Blobcat} results are comparable. As expected, both algorithms were found to perform very well to detect 
bright and faint compact sources, including blended sources, but they are biased, by design, against extended sources. A 5$\sigma$ threshold was considered for source detection 
with the Wavelet method on two different scales $J$=5, 6. In these conditions, most of the extended bright sources present in the fields can be detected. Fainter features, such as 
parts of the supernova remnants or diffuse regions cannot be well detected, at least at the specified significance level. 

The Chan-Vese algorithm was tested over the residual image under different choices of parameters and using a simple circular level-set as initial contour. A pre-smoothing stage 
is applied to the input residual image. Contours surrounding areas of negative excesses with respect to the background level were removed from the set of final detected contours. As 
it can be seen, the extended source features missed by the other algorithms can be extracted with high accuracy compared to the human segmentation. Some imaging artefacts are also detected 
along with real sources even with the optimal choice of the Chan-Vese parameters. Overall, the Chan-Vese method was found to outperforms the other three tested algorithms in 
fully detecting extended objects.

In Fig.~\ref{SourceFluxComparisonFig} we compare the integrated flux of the extended sources present in the four fields A-D estimated with three different methods 
(\tool{}: black dots, Chan-Vese: red squares, Wavelet method at scale $J$=5: blue triangles) as a function of the flux estimated using the human-driven segmentation. A total of 30 source 
candidates were identified, hereafter denoted as the "reference set". Data are reported in the plot for each algorithm in case of source identification and cross-match found with 
the reference set. As it can be seen, the estimated fluxes closely follow the reference, the observed spread in flux being regarded as a measure of the source reconstruction 
accuracy contribution to the total flux uncertainty. Overall, better results are obtained with the \tool{} and Chan-Vese algorithm, which are able to detect fainter sources with 
respect to the Wavelet method and achieve a better accuracy in flux estimation.

We are aware that we have not exhausted the list of all possible algorithms for extended source extraction and that a deeper tuning is needed for the three tested algorithms before 
drawing firm conclusions on their suitability for our goals. For instance, a more refined initialization strategy is desired in the Chan-Vese method together with a finer 
exploration of the parameter space. Moreover, it is known that the two-level assumption (foreground/background) at the basis of the standard Chan-Vese algorithm may not be 
accurate to scenarios in which a large variation of intensity levels is present. New active contours algorithms \citep{VeseChan2002,Yang2013}, overcoming some of the standard Chan-Vese 
limitations, appeared recently in the literature and could be worthy of consideration. However, we expect that none of the methods will perform accurately 
over all the presented images and that a combination of different techniques is probably required at the very end. That motivated the development of a completely 
different approach reported in this paper. 

\section{Summary}
We described in this paper a new algorithm for the detection of extended sources in radio maps, designed for the SCORPIO project and for next-generation 
radio surveys. The algorithm was tested with real radio data observed in the SCORPIO and Molonglo surveys and compared with existing algorithms. The achieved performances are 
found comparable or even superior to other approaches followed in the literature. The novel points introduced are:
\begin{itemize}
 \item a new procedure for computing the background in presence of extended emission;
 \item an efficient filter to enhance diffuse emission, based on compact bright source removal, smoothing and saliency estimation;
 \item a flexible framework providing rich information for post-processing analysis and relaxing some of the limiting requirements used for compact source 
 detection (e.g. pixel adjacency)
\end{itemize}
The results obtained with real data are promising and motivate further work both on the data side and on the algorithm side.

For this purpose, a new release of the SCORPIO map, with improved cleaning procedure and data flagging applied, is in progress. Preliminary results on 
the studied fields show that many of the artefacts present in the first data release are now properly removed. Further, a campaign of single-dish measurement in the SCORPIO field is 
already scheduled to improve the map response to extended objects beyond the limits of the ATCA telescope. Source finding will therefore largely benefit from these improved maps.

At the same time, simulation activities were started with the aim of generating extended source mock scenarios with ground truth available at pixel level to study the achieved 
source detection efficiency and contamination rate with realistic noise conditions.

We are currently working on possible significant improvements also on the algorithm side, both at code and method level. Among these, improving saliency estimation and resolution has 
become an active field of development in recent works, see \citet{Perazzi2012,Cheng2014,Borji2014,Shi2015}. A proper combination of different algorithms 
could be a viable solution to decrease the spurious detection rate. Suitable criteria for combining nearby candidate sources is another aspect to be investigated in detail.

The current algorithm implementation is not optimized for large maps, e.g. the full SCORPIO or expected ASKAP fields, as it still requires large computation 
time, e.g. from few to $\sim$15-20 minutes depending on image size, and memory requirements even on a single field, mainly related to the superpixel similarity matrixes. 
A new optimized version, also designed for parallel and/or distributed processing is therefore planned to be realized, possibly compliant with ASKAP EMU software pipeline 
requirements in terms of input/output products to be supported, employed technologies and processing strategies \citep{Cornwell2011,Chapman2014}.





\bsp	
\label{lastpage}
\end{document}